\documentclass[aps,prl,twocolumn,superscriptaddress,notitlepage,nofootinbib,longbibliography,10pt]{revtex4-2}
\usepackage[colorlinks=true, allcolors=blue, citecolor=blue]{hyperref}
\usepackage{mathrsfs}
\usepackage{epsfig}
\usepackage{graphicx}
\usepackage{amsfonts}
\usepackage[figuresright]{rotating}
\usepackage{amssymb}
\usepackage{amsmath}
\usepackage{dcolumn}
\usepackage{bm}
\usepackage[dvipsnames]{xcolor}
\usepackage{braket}
\usepackage{units}
\usepackage{xspace}
\usepackage{bbold}
\usepackage{orcidlink}
\usepackage{tabularx}
\usepackage{adjustbox}
\usepackage{comment}
\usepackage[]{hyperref}

\newcommand{\JSU}{Department of Physics, Jiangsu University, Zhenjiang, 212013, China}
\newcommand{\NU}{School of Physics and Technology, Nantong University, Nantong, 226019, China}
\newcommand{\COMSATS}{Department of Physics, COMSATS University Islamabad, Islamabad Campus, 45550, Pakistan}
\begin{document}
\title{Sub-shot-noise sensitivity via superpositions of two deformed kitten states}

\author{Naeem Akhtar}
%\email{naeemakhtar166067@gmail.com}
\affiliation{\JSU}
\author{Xiaosen Yang}
\email{yangxs@ujs.edu.cn}
\affiliation{\JSU}
\author{Jia-Xin Peng}
\email{JiaXinPeng@ntu.edu.cn}
\affiliation{\NU}
\author{Inaam Ul Haq}
\affiliation{\COMSATS}
\author{Yuee Xie}
\affiliation{\JSU}
\author{Yuanping Chen}
\email{chenyp@ujs.edu.cn}
\affiliation{\JSU}
\date{\today}
\begin{abstract}

In the present work we explore nonclassical effects in the phase space of two superposed kitten states induced by photon addition and subtraction operations applied in different sequences. We investigate two scenarios: In the first, photon addition is applied to the state, followed by photon subtraction, while in the second, the order of operations is reversed. We demonstrate that applying multiphoton operations to the state results in notable nearly isotropic sub-Planck structures, with the characteristics of these structures being influenced by the photon addition and subtraction. Increasing the number of added photons compresses the sub-Planck structures in both cases. Photon subtraction, however, has the opposite effect on the sub-Planck structures in the first case and no effect in the second. Furthermore, we observe that the optimal choices of multiphoton operations lead to improved isotropy of sub-Planck structures in our cases. The presence of the sub-Planck structures in our states leads to improved sensitivity to displacements, exceeding the standard quantum limit, as verified across all the depicted scenarios.

\end{abstract}
\maketitle
\section{Introduction}
Coherent states were first introduced by Schr\"{o}dinger \cite{Sch26,MN09}, and the concept was further developed in quantum optics by Glauber~\cite{Gla63}. Quantum superposition phenomena have been intensively investigated within the framework of the harmonic oscillator, leading to the creation of intriguing nonclassical states~\cite{Gerry05book,BUZEK19951}. The nonclassical nature of quantum states is revealed through nonclassical phase-space features, often visualized using the Wigner function~\cite{Wig32,Sch01}. Nonclassical states are considered valuable resources for continuous variable quantum information processing, particularly for sensing and metrology applications~\cite{Joo2011,pan2014,mitchell2004,johnson2017ultrafast,Facon2016}.\;A critical factor in determining the utility of a quantum state for such applications is the degree to which it becomes distinguishable from the initial state after a small displacement, which is influenced by the smallest phase-space feature of a quantum state and measures sensitivity to displacement (perturbation)~\cite{Zurek2001,Audenaert14,DIEKS1988303}.

The phase-space volume occupied by a coherent state adheres to the Planck scale $\hbar$~\cite{CarlosNB15}, while also preserving its sensitivity to displacements at this standard limit~\cite{leonhardt1997measuring, Robertson1929}. Developing quantum states with finer phase-space features is essential for enhancing their capacity to measure smaller scale displacements, and quantum states with finer phase-space features are highly valuable for advancing quantum measurement techniques and improving the performance of quantum technologies~\cite{Braunstein2005}, which specifically translated to sensing platforms based on superconducting circuits~\cite{PhysRevLett.122.080502,blais2020quantum,campagne2020quantum}, trapped ion systems~\cite{PhysRevLett.116.140402,Bruzewicz2019}, and magnon sensing~\cite{PhysRevLett.125.117701,Dany2020}, to name a few. Nonclassical states are also crucial to detecting weak forces (tidal gravitational forces)~\cite{nemoto2003quantummetrology,Toscano06,Eff4,PhysRevX.9.021023,Gilchrist_2004}. In these contexts, minimal uncertainty states~\cite{CarlosNB15}, such as coherent states, define the classical limit for force detection. This limit can be surpassed by quantum states exhibiting finer phase-space features~\cite{nemoto2003quantummetrology, Maccone2004}, with certain chaotic quantum systems demonstrating states whose phase-space volumes are significantly smaller than those of coherent states~\cite{Zurek2001, Jordan2001}.

The effectiveness of weak-force detection techniques appears to be directly impacted by the degree of phase-space feature fineness in quantum states. For example, a macroscopic cat state (superpositions of two distinct coherent states)~\cite{Mil86,YS86} exhibits finer phase-space structures than coherent states; that is, such features are constrained along a specific direction in the phase space, thus holding the capacity to detect the directional weak forces~\cite{Toscano06,Eff4,nemoto2003quantummetrology}. The notion of coherent-state superpositions has been evolved to a generalized form of macroscopic cat states~\cite{PhysRevA.99.063813,Howard2019,Zurek2001,johnson2017ultrafast}, and
quantum compasses~\cite{Toscano06, DODONOV2016296, Eff1, Eff2, Eff3, Eff5, Eff7, Eff8, Eff9, Eff10, Naeem2021, Naeem2022, Naeem2023, Naeem2024, Howard2019, Atharva2023,Zurek2001} are particular examples of such generic superpositions and are notable in that they exhibit improved phase-space characteristics compared to their precursors, namely coherent states and cat states. In particular, it has been found that these states contain sub-Planck scale features (dimension below the Planck scale)~\cite{Zurek2001}, which are extremely fragile against perturbation and exhibit sensitivity to displacements greater than the standard quantum limit, making them a potential candidate for quantum metrology applications~\cite{Toscano06,Eff4,Park2023quantumrabi}. Sub-Planck features have been found in entangled Schr\"{o}dinger cat states created in a two-cavity setup~\cite{Batin2024}, extending the notion of sub-Planck structures to higher-dimensional cat states.

Sub-Planck features, also known as blind spots~\cite{Zambrano_2009}, are constrained along all phase-space directions, allowing for the simultaneous measurement of conjugate variables (for a specific case, position, and momentum) with maximum precision~\cite{Toscano06}. These microscopic characteristics have been further extended to precisely determine the displacement of a mechanical oscillator or microwave field in any arbitrary direction within phase space~\cite{Eff4,Toscano06}, making them superior to the cat state features. In addition to such specific cases, nonclassical states~\cite{PhysRevA.95.012305,RevModPhys.90.035006}, particularly compass states~\cite{Park2023quantumrabi}, have also played a crucial role as probes for enhanced phase and displacement estimation. In quantum communication channels, the robustness of the information carrier to noise is crucial~\cite{Teklu_2015,PhysRevAccelBeams24,Teklu19,Trapani2015}, and the use of finer-scale characteristics of the Wigner function has been shown to improve the effectiveness of the communication protocol, resulting in high-fidelity continuous-variable teleportation~\cite{SCOTT20082685}. The effectiveness of finer phase-space characteristics at the sub-Planck level is crucial in continuous-variable quantum information ranging from quantum estimation to communication, and this work particularly focuses on the sub-Planck structures that are closely tied to the compasslike state of the harmonic oscillator.

Both theoretical~\cite{Prop1,Prop2,Prop4,Prop5,choudhury2011proposal} and experimental~\cite{Exp1,Exp2,Exp3,Exp4} approaches have been employed to generate catlike states. Multiphoton operations applied to a quantum state provide innovative techniques for controlling fundamental nonclassical characteristics~\cite{PhysRevA.86.012328, Wenger, Dakna1997, Tang201586, Alexei2006, Neergaard2006, Wang11,Barnett2018, Guerrini, Zavatta_2008,Chen2024, Lund2024, Takase2021, PhysRevA.94.063830}. For example, it has been shown that photon subtraction (or addition) from ordinary squeezed-vacuum states may develop quantum states similar to those of cat states~\cite{Dakna1997, Neergaard2006, Chen2024, Alexei2006}. In this work, we consider a four-headed kitten state, which is the superposition of two kitten states deficient in nonclassical phase-space attributes. Photon addition and subtraction operations are applied on this multi-headed kitten state in different orders to build new quantum states, which exhibit intriguing phase-space characteristics similar to a compass state~\cite{Zurek2001}. In our first case, we apply photon addition followed by photon subtraction to the state, while in the second case, the order of operations is reversed, with photon subtraction applied first and then photon addition.

\textbf{\small{Contribution}.} Our investigation utilizes phase-space formalism~\cite{Sch01}, incorporating Wigner function analysis and photon number distributions, and we compare the original state to its deformed variants acquired from multiphoton operations by performing an overlap function between the states. This comparison investigates the ranges of parameters where the effects of multiphoton operations are highest and gives the ideal parameter selection, and we work in the region where our proposed states tend to be distinct from the original state. We also analyze displacement sensitivity for each case and present a thorough discussion of the physical significance and implications. Specifically, we show that multiphoton operations applied on the four-headed kitten state lead to quantum states with refined phase-space characteristics, and interestingly, our proposed cases also exhibit sub-Planck phase-space features whose occurrence is now affiliated with the number of added and subtracted photons. Furthermore, we show that our proposed quantum states exhibit sensitivity to phase-space displacements greater than the standard quantum limit, making them a promising choice for quantum sensing applications. 
 
 Our observations indicate that as the number of added photons increases, finer sub-Planck structures develop in the phase space of our indicated cases. In contrast, in the first case, when addition is applied prior to subtraction, corresponding photon subtraction in this case leads the sub-Planck structures to grow in size, resulting in the loss of the sub-Planck scale characteristic. In the second scenario of subtraction followed by addition, the photon subtraction operation is ineffective. In our cases, an improvement in the isotropy (directional invariance in phase space) of the sub-Planck structures is observed for certain photon subtraction choices. These findings align with the behavior observed in the sensitivity of our quantum states.

\textbf{\small{Organization}.} \S\ref{sec:quad_sens} discusses the impact of phase-space features, particularly sub-Planck structures, on sensitivity to displacement. This analysis is carried out by comparing cat states and compass states in terms of their practical applications in estimation schemes. \S\ref{sec:photon_varied} discusses the deformed version of quantum states; specifically, our focus is on the photon-varied version of coherent states and their derivatives, such as multicomponent kitten states, with a particular focus on their sub-Planck phase-space characteristics and variations over involved parameters and multiphoton operations. \S\ref{subsec:sens_var} analyzes how the sub-Planckian features in our cases have the impact on the enhancement in phase-space sensitivity to displacements. \S\ref{sec:outlook} provides a compact summary of our findings and \S\ref{sec:conc} provides the main conclusions of our results.

\begin{figure*}[t]
\includegraphics[width=1.03\textwidth]{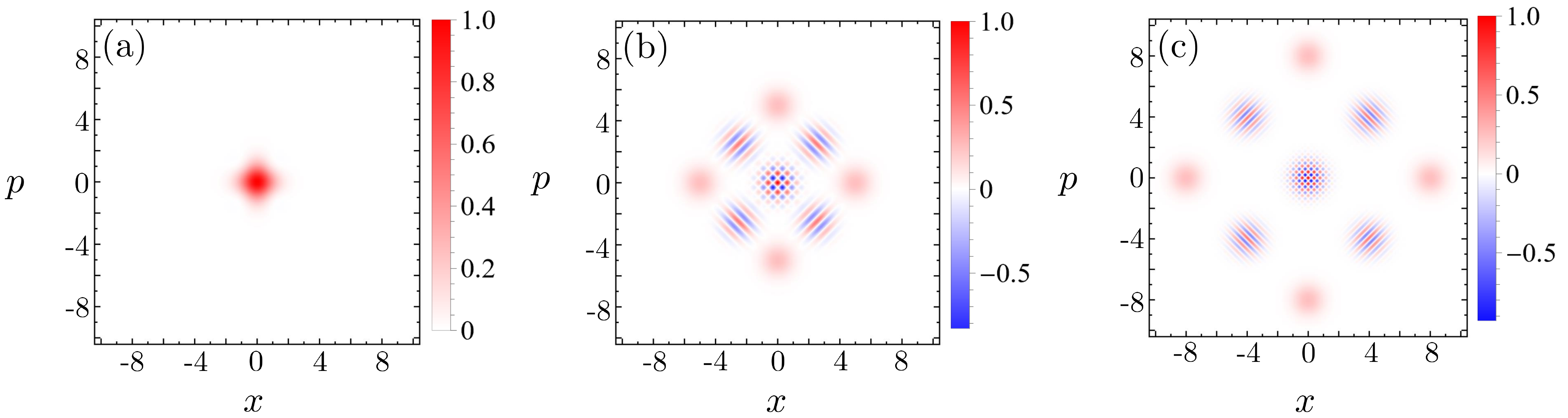}
\caption{Wigner function of the compass state. for (a)~$c_0=1$, (b)~$c_0=5$, and (c)~$c_0=8$.}
\label{fig:fig1}
\end{figure*}

\section{Concepts and perspectives}\label{sec:quad_sens}

The quantum uncertainty principle for position and momentum, expressed through the commutation relation $[\hat{x},\hat{p}]:=\text{i}\hbar$, with $\hat{x}$ and $\hat{p}$ the position and momentum operators, respectively, imposes constraints on the dimension of a phase-space structure. Specifically, it dictates that the product of the uncertainties in position ($\Delta x$) and momentum ($\Delta p$) satisfies $\Delta x \Delta p \geq \nicefrac{\hbar}{2}$~\cite{leonhardt1997measuring, Robertson1929}. This implies that the dimension of a phase-space feature is constrained by this standard limit; if a phase-space feature has dimensions below this threshold, it is considered nonphysical. This general assumption was challenged by Zurek~\cite{Zurek2001}, who demonstrated that the sub-Planck structures in the compass state are significant; that is, these spotty features have been found to be crucial in enhancing sensitivity to displacements, and this improvement in sensitivity is directly dedicated to the sub-Planck scale structures in the state. In this section we primarily focus on reviewing the key concept of sub-Planck structures and their critical role in enhancing sensitivity to displacements.

\subsection{Phase space and sensitivity}\label{subsec:main_theory}

A Schr\"{o}dinger coherent state can be expressed as a displaced vacuum state $\ket{\alpha}:=\hat{D}(\alpha)\ket{0}$, where $\hat{D}(\alpha):=\exp(\alpha\hat{a}^\dagger-\alpha^*\hat{a})$ is the displacement operator~\cite{Gaz09}, with $\alpha \in \mathbb{C}$ and $\hat{a}$ ($\hat{a}^{\dagger}$) the annihilation (creation) operator. Coherent states have properties that mirror classical states~\cite{CarlosNB15}, but their superposition may exhibit nonclassical features due to the interference phenomena~\cite{Weinbub2018}.
The Wigner function denoted by $W_{\hat{\rho}}(\bm{\beta})$, with $\bm{\beta}:=(x,p)^{\top}$, constitutes the phase space of a quantum state $\hat{\rho}$~\cite{Sch01}, where $x$ and $p$ are the position and momentum pairs, respectively.

The Wigner negativity is one of the primary tools for analyzing the nonclassical characteristics of a state~\cite{Sch01,PhysRevA.97.013840,Guise1999,Nunes2019,teklu2015nonlinearity}. 
Mathematically, the Wigner function can be expressed as~\cite{FAN1987303}
\begin{align}\label{eq:Wig1}
W_{\hat{\rho}}(\bm{\beta}):=\frac{\text{e}^{2|\beta|^2}}{\pi^2}\int d^2\gamma\braket{-\gamma|\hat{\rho}|\gamma}\text{e}^{-2\left(\beta^*\gamma-\beta\gamma^*\right)}.
\end{align}
Note that dimensionless versions of the position and momentum operators are employed throughout this work.

The Schr\"{o}dinger cat state is the superposition of two distinguishable coherent states, and one of the simplest examples of such states is the even cat state~\cite{DODONOV1974597}, which is defined as
\begin{align}\label{eq:evencat}
 \ket{\psi}:=\frac{1}{(2+2\text{e}^{-2|\alpha|^{2}})^{\nicefrac{1}{2}}}\left(\ket{\alpha}+\ket{-\alpha}\right).
\end{align}
A cat state is obtained when the parameter $\alpha$ is high enough to maintain the orthogonality criterion $\braket{\alpha|-\alpha}\approx 0$. If the constituent coherent states in this superposition are not distinguishable, the resulting states are known as Schr\"{o}dinger kitten states~\cite{Alexei2006}. This generic form of the cat state with the choices $\pm\alpha=\nicefrac{\pm c_0}{\sqrt{2}}$ ($c_0 \in \mathbb{R}$) represents a horizontal cat state aligned along the position axis in the phase space. 

We now discuss the concept of sensitivity to phase-space displacement. For a pure quantum state $\ket{\psi}$, this sensitivity can be mathematically determined by evaluating the overlap function between a quantum state and its slightly displaced version~\cite{Zurek2001,Audenaert14,DIEKS1988303,Toscano06}. This involves calculating how much one state resembles another when displaced in phase space, which provides insight into how precisely the quantum state can detect or respond to changes in its phase-space configuration. Mathematically, this sensitivity can be determined by using~\cite{Audenaert14}

\begin{align}\label{eq:sens1}
 S_{\ket{\psi}}(\delta):=\int \frac{d^2{\beta}}{\pi} W_{\ket{\psi}}(\beta)W_{\ket{\psi^{\prime}}}(\beta)=\left|\braket{\psi|\psi^{\prime}}\right|^2
\end{align}
with $\ket{\psi^{\prime}}:=\hat{D}(\delta)\ket{\psi}$. If $S_{\ket{\psi}}(\delta)=0$, then a state and its displaced counterpart are orthogonal for the displacement $\delta$. Here $S_{\ket{\psi}}(\delta)$ is the overlap, with $\delta:=(\delta x,\delta p)^{\top}$, where $\delta x$ and $\delta p$ are values of the displacements applied in the phase space along position ($x$) and momentum ($p$) axis, respectively. The infinitesimal perturbation $\delta$, which makes the perturbed state quasiorthogonal to the initial state, provides information on the sensitivity to displacements. Smaller values of $\delta$ indicate greater sensitivity to displacements.

\begin{figure}
\includegraphics[width=0.52\textwidth]{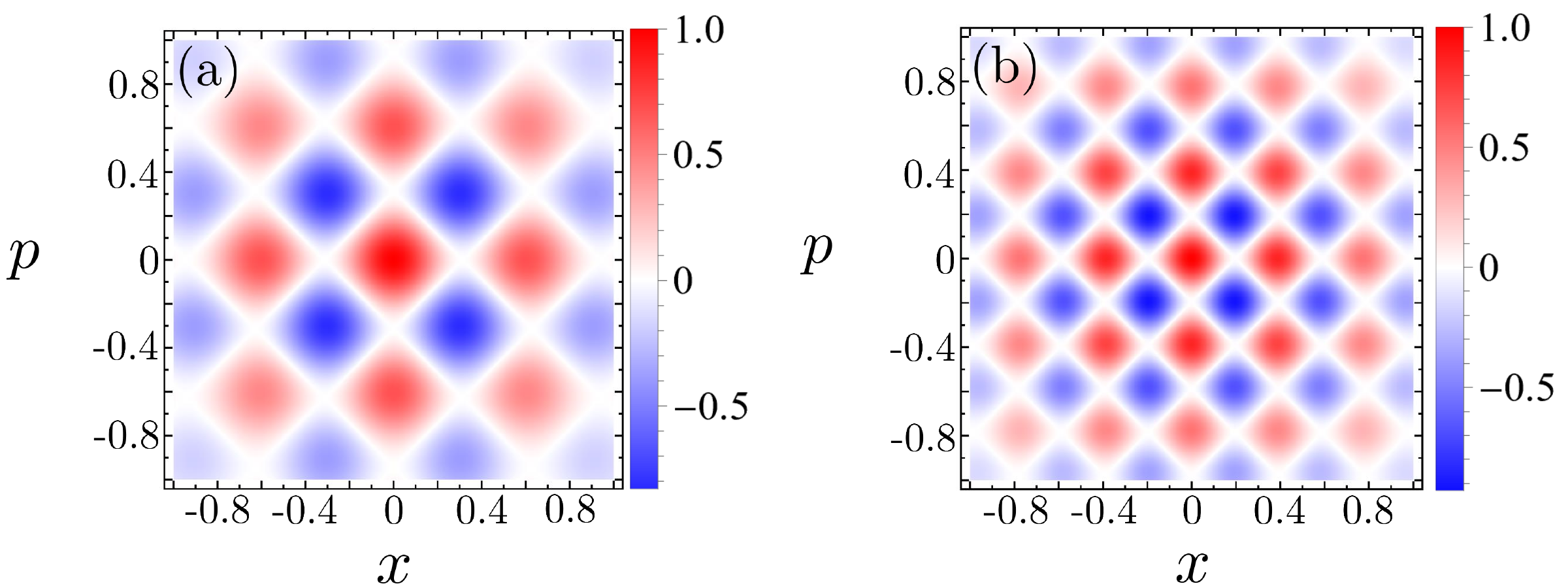}
\caption{Central interference of the compass state for (a)~$c_0=5$ and (b)~$c_0=8$.}
\label{fig:fig2}
\end{figure}

The Wigner function of a coherent state is denoted as

\begin{align}
W_{\ket{\alpha}}(\bm{\beta}):=\text{e}^{-2(\alpha-\beta)(\alpha^*-\beta^*)},
\label{eq:wig_coh}
\end{align}
which exhibits a Gaussian form; hence, coherent states are types of Gaussian quantum states~\cite{CarlosNB15}, as also evident by the corresponding Wigner function of the coherent state shown in Fig.~\ref{fig:figapp}(a).
Moreover, the phase-space structure of a coherent state follows the minimal limit set by the uncertainty principle, often referred to as the Planck action in phase space. This implies that the phase-space size of a coherent state sets the minimal norm, and the sub-Planck structure is below this limit and can be limited as much as desired by varying the controlling parameter, whereas all of these characteristics are missing in the coherent~\cite{Naeem2021} (the corresponding discussion about the Planck scale is provided in the Appendix~\ref{appendix:appendixA}). The sensitivity of a coherent state to displacements in phase space is obtained as
\begin{equation}\label{eq:sens_coh}
S_{\ket{\alpha}}(\delta):=\text{e}^{-|\delta|^2},
\end{equation}
which tends to zero for the displacement $|\delta| > 1$, implying that the sensitivity of a coherent state falls at the standard limits. The phase-space structure of a coherent state and its sensitivity adhere precisely to the standard quantum-mechanical limits, that is, the coherent state achieves the theoretical minimum uncertainty allowed by the Heisenberg uncertainty principle, reflecting the optimal balance between precision in position and momentum measurements. Consequently, in our analysis, we evaluate each example by comparing it against these established norms. This involves assessing how each example measures up to the theoretical benchmarks and standard limits, allowing us to understand their relative performance and behavior in relation to these reference points.

\subsection{Direction-oriented sensitivity enhancement}\label{subsec:cat_sens}

The Wigner function of a macroscopic horizontal cat state typically manifests as two distinct Gaussian peaks in phase space with a central oscillatory interference pattern directed along the momentum axis, where each Gaussian peak corresponds to a coherent state~\cite{Thekkadath2020engineering, Lee15}. Cat states do not exhibit sub-Planck features as their interference phase-space features are not limited in all directions of phase space~\cite{Naeem2021}. Figures~\ref{fig:figapp}(b) and \ref{fig:figapp}(c) shown in Appendix \ref{appendix:appendixA} illustrate the central phase-space parts of a horizontal cat state with smaller (kitten state) and larger (macroscopic cat) amplitudes $c_0$, respectively. Mathematically, the Wigner function of the horizontal cat state is defined as
\begin{align}\label{eq:cat_wig}
    W_{\ket{\psi}}(\beta):=\frac{1}{[2+2\text{exp}(-c_0^{2})]}\sum^2_{i,j=1}W_{\ket{\alpha_i}\bra{\alpha_j}}(\beta),
\end{align}
where
\begin{align}\label{eq:gen1_wig}
W_{\ket{\alpha_i}\bra{\alpha_j}}(\beta):=&\nonumber G_{\alpha_i,\alpha_j}\exp\big[-\alpha_i\alpha^*_j-2\big(|\beta|^2-\alpha^*_j\beta\\ &-\alpha_i\beta^*\big)\big],
\end{align}
with
\begin{align}
    G_{k,l}:=\exp\bigg[-\frac{1}{2}\left(|k|^2+|l|^2\right)\bigg],
\end{align}
setting $\alpha_1 = c_0$ and $\alpha_2 = -c_0$ in Eq.~(\ref{eq:cat_wig}), with $c_0 = 1$ representing the horizontal kitten state and $c_0 = 5$ representing its larger version. Figures~\ref{fig:figapp}(b) and \ref{fig:figapp}(c) in the Appendix~\ref{appendix:appendixA} clearly illustrate the difference in phase-space features between the smaller and larger cats, respectively. In comparison to its kitten counterpart and the coherent state, the larger cat state appears to have more fine-grained phase-space features. For example, the central phase-space feature, shown in Fig.~\ref{fig:figapp}(c), is constrained along the $p$ direction in the phase space, but it has the same extension along the $x$ direction as those of coherent states, which is highlighted by plotting corresponding zeros indicated with black curves around the patches. This suggests a directional dependence on the finer resolution of the corresponding interference features, which also affects the sensitivity to displacements in the same way. The sensitivity to displacement for a cat state is obtained as
\begin{align}
S_{\ket{\psi}}(\delta):=\left|\sum^2_{i,j=1}O_{\ket{\alpha_i}\bra{\alpha_j}}(\delta)\right|^2,
\end{align}
where
\begin{align}
O_{\ket{\alpha_i}\bra{\alpha_j}}(\delta):=G_{\alpha_i,\alpha_j}\exp\left[\alpha^*_i\alpha_j+\alpha^*_i\delta-\alpha_j \delta^*-\frac{|\delta|^2}{2}\right].
\end{align}
The overlap for $c_0 = 5$ is shown in Fig.~\ref{fig:figapp}(i), and it can be verified that the overlap vanishes for $|\delta| < 1$ along the momentum axis. Along the position axis in phase space, however, $|\delta| > 1$ is needed to make the overlap zero, which aligns with that of a coherent state in the same direction, suggesting that for a cat state, the sensitivity to displacement is heightened along specific phase-space directions. In this case, it is confined along the momentum direction, as shown by the white curves representing the zeros of the overlap function. Additionally, it can be shown that the central structure of the overlap function for the horizontal cat state may contract further along the momentum axis as $c_0$ increases, indicating that the sensitivity to displacement improves as $c_0$ grows.

Another illustration is the Wigner phase-space features of squeezed-vacuum states, which are limited along the direction of squeezing, implying that in these directions the sensitivity to displacement can be less than that of a coherent state. This shows that squeezed-vaccum states may provide enhanced sensitivity compared to standard coherent states, making them useful in quantum metrological applications~\cite{drummond2004quantum,Maccone2004}.

\begin{figure}[t]
\includegraphics[width=0.5\textwidth]{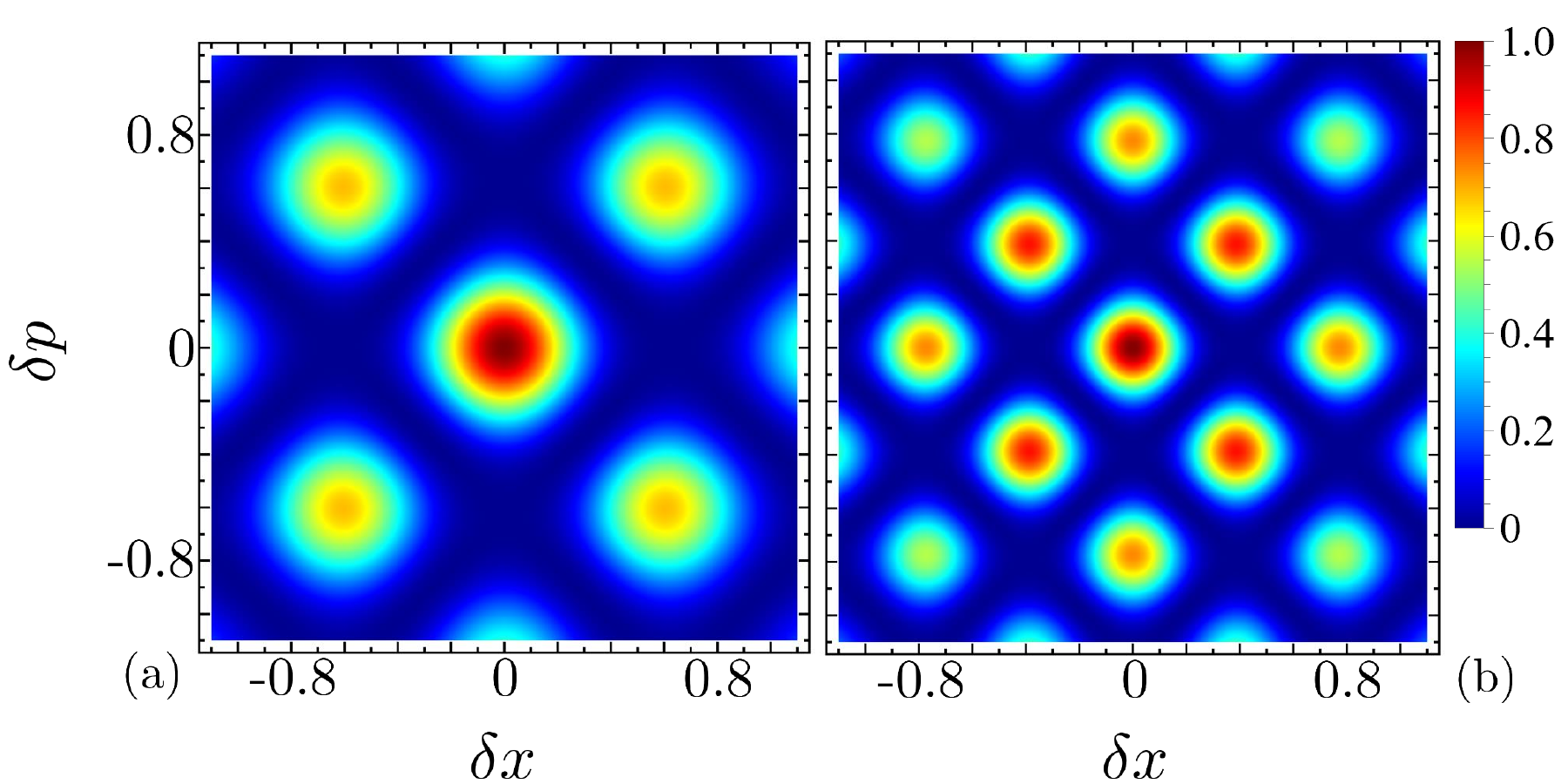}
\caption{The overlap between the compass state and its displaced versions, represented by $S_{\diamondsuit}(\delta)$, quantifies the corresponding sensitivities over given parameters chosen with (a)~$c_0=5$ and (b)~$c_0=8$. The intensity plots are normalized to unity for each case.}
\label{fig:fig3}
\end{figure}

\subsection{Sub-Planckian sensitivity}\label{subsec:compass}

We now include the example of the Zurek compass state~\cite{Zurek2001}, which is recognized as a superposition of coherent states given by $\alpha_1 = \nicefrac{c_0}{\sqrt{2}}$, $\alpha_2 =\nicefrac{-c_0}{\sqrt{2}}$, $\alpha_3 = \nicefrac{\text{i}c_0}{\sqrt{2}}$, and $\alpha_4 = \nicefrac{-\text{i}c_0}{\sqrt{2}}$, with $c_0 \in \mathbb{R}^+$. This superposition can also be interpreted as the superposition of two cat states, or equivalently as the superposition of four coherent states and is defined as

\begin{align}\label{eq:compass0_state}
  \ket{{\diamondsuit}}:=N^{\nicefrac{-1}{2}}_{{\diamondsuit}}\sum^4_{i=1}\ket{\alpha_i},
\end{align}
where
\begin{align}
  N_{\diamondsuit}=\sum^4_{i,j=1}G_{\alpha_i,\alpha_j}\text{e}^{\alpha^*_i\alpha_j}
\end{align}
represents the normalization coefficient.

The Wigner function of the compass state $\ket{\diamondsuit}$ is obtained as
\begin{align}
W_{\ket{\diamondsuit}}(\beta)=\frac{1}{N_{\diamondsuit}}\sum^4_{i,j=1}W_{\ket{\alpha_i}\bra{\alpha_j}}(\beta).
\end{align}
We present $W_{\ket{\diamondsuit}}(\beta)$ in Fig.~\ref{fig:fig1} for a few $c_0$ values. As depicted in Fig.~\ref{fig:fig1}(a), for $c_0 = 1$, the four coherent states in the compass state cannot be distinguished individually; hence, the corresponding Wigner distribution shows a central positive peak. This scenario represents the four-component kitten state; alternatively, it may also be referred to as a four-headed kitten state~\cite{Lee15}. However$c_0=5$, these four coherent states are now well separated and appear as four Gaussian lobes in the phase space, and the interference pattern is now pronounced in the phase space, as shown in Fig.~(\ref{fig:fig1})(b). Figure~\ref{fig:fig1}(c) shows that increasing the macroscopic parameter $c_0=8$ causes coherent states to be pushed further away from the phase-space origin, resulting in enhanced negative regions in the intensity plot. 
Figure~\ref{fig:fig2} illustrates the central interference pattern of cases presented in Figs.~\ref{fig:fig1}(b) and \ref{fig:fig1}(c), where the phase-space features are arranged in a tiled format,  and each tile in this pattern has an extension considerably smaller than that of a coherent state, indicating that these structures are at the sub-Planck scale, and as $c_0$ increases, the size of the sub-Planck features reduces. Furthermore, these tilelike sub-Planck structures represent their anisotropic version, as the corresponding structures are not uniformly constrained in the phase space. The discussion of the sub-Planckian scale, as well as the comparison between isotropic and anisotropic phase-space features, can be further explored in the Appendix~\ref{appendix:appendixA}.

The overlap between the compass state $\ket{\diamondsuit}$ and its displaced version $\hat{D}(\delta)\ket{\diamondsuit}$ reflects the sensitivity to displacement in phase space
\begin{align}
    S_{\diamondsuit}(\delta)=\left|\sum^4_{i,j=1}O_{\ket{\alpha_i}\bra{\alpha_j}}(\delta)\right|^2.
\end{align}
Figures \ref{fig:fig3}(a) and \ref{fig:fig3}(b) demonstrate that the overlap function $S_{\diamondsuit}(\delta)$ is zero for $|\delta|< 1$ along any direction in the phase space, with $c_0 = 5$ and $c_0 = 8$ in the corresponding situations involving the sub-Planck structures. This means that the sensitivity to displacement for this compass state is increased when compared to coherent states as well as when compared to cat states because the corresponding enhancement occurs along all phase-space directions. Furthermore, compared to the coherent state, the overlap function is now dependent on $c_0$. Increasing this parameter causes the overlap function to be zero for smaller values of $|\delta|$, which leads to a greater enhancement in the corresponding sensitivity. The enhanced sensitivity achieved by this compass state exhibits anisotropy. This becomes apparent when observing the central patches in Figs.~\ref{fig:fig3}(a) and \ref{fig:fig3}(b), where the structures appear tilelike. Further evidence of this anisotropy can be clearly observed in Fig.~\ref{fig:figapp}(j) of Appendix~\ref{appendix:appendixA}, where the directional dependence is apparent. This indicates that the sensitivity enhancement does not occur uniformly in all directions but rather is more pronounced in certain orientations in phase space.

\subsection{Measurement strategies}

In this section we mainly elaborate how the detection of weak forces is influenced in a detection scheme that utilizes nonclassical states, with particular emphasis on cat states versus compass states. The precision of quantum parameter estimation is influenced by the energy resources employed during the measurement, such as the average photon number. As indicated in \S\ref{subsec:main_theory}, the sensitivity of a coherent state $\ket{\alpha}$ is independent of the specific value of $\alpha$ [Eq.~(\ref{eq:sens_coh})], which is proportional to the average photon number by $|\alpha|^2$. This implies that increasing the average photon number is irrelevant to enhancing the sensitivity of a coherent state to displacements, as it is limited by shot noise from vacuum fluctuations~\cite{CarlosNB15}, a limitation that can be overcome using quantum effects such as superposition and squeezing~\cite{Maccone2004}. For the same average photon number, cat states outperform coherent states. However, compass states show superior sensitivity to displacements compared to both coherent and cat states [as also indicated in Figs.~\ref{fig:figapp}(i) and \ref{fig:figapp}(j)], enhancing precision in weak-force detection~\cite{Toscano06, Eff4}. In the following, we precisely elaborate on how these detection schemes work.

The concept of weak force detection can be easily understood by considering that a signal one wish to measure is linearly coupled to a harmonic oscillator, causing the oscillator to experience a displacement proportional to the strength of the signal. The sensitivity of the oscillator to these displacements determines the strength of the signal, meaning that quantum states with finer phase-space features (greater sensitivity) are capable of detecting weaker signals. From Eq.~(\ref{eq:sens_coh}), it is clear that coherent states can measure displacements of order $|\delta| \sim 1$, setting the standard classical limit for measurement schemes. However, for horizontal cat states with comparatively finer features, as observed in Fig.~\ref{fig:figapp}(i), the displacement along the $x$ axis behaves similarly to a coherent state, as in this direction the extension of the patch is also the same as that of the coherent state, while along the momentum axis, the central patch exhibits finer resolution compared to the coherent state, and in this direction the sensitivity is inversely proportional to $c_0$, which, for $c_0 \gg 1$, exceeds the classical limit, also called the Heisenberg limit of the sensitivity. The sensitivity to displacement for compass states scales similarly to that of cat states. However, as shown in Fig.~\ref{fig:figapp}(i), unlike cat states, this enhancement is independent of orientation in phase space, allowing compasses to detect weak forces regardless of their direction.

Cat and compass states have been utilized to detect tiny displacements and rotations in phase space, achieving the Heisenberg limit of the sensitivity, especially in cavity systems and systems based on ion traps~\cite{PhysRevA.45.5193, Schneider1998, Eff4, nemoto2003quantummetrology}. For example, in the case of detecting weak forces by using cavity QED and
ion trap systems~\cite{Eff4,Toscano06}, it has been demonstrated that quantum states with coarser phase-space structures, such as cat states, have disadvantages compared to those with finer phase-space features, like sub-Planck structures in such measurement schemes. More specifically, the sensitivity of cat states gradually decreases as the direction of the perturbing force deviates from the axis orthogonal to the line joining the two coherent states.

In metrological applications, achieving simultaneous accuracy in parameters that typically have a trade-off in their measurement precision, such as conjugate variables or other related parameters~\cite{PhysRevResearch.5.013185}, is a critical challenge. Compasses, on the other hand, have the capacity to get the highest precision in the measurement of conjugate variables at the same level, which has been analyzed in the aforementioned cases ~\cite{Eff4,Toscano06}. Additionally, a conventional interferometer can provide an unambiguous estimation of displacement in a known direction in phase space~\cite{Suda2001,Garbe2020,McCormick_2019}. A hybrid oscillator-qubit interferometric setup using compass state as its local state has demonstrated advantages over the conventional interferometer, as this protocol achieved the unambiguous estimation of phase-space displacements in an unknown direction of a mechanical oscillator~\cite{Park2023quantumrabi}.

In summary, quantum compasses have proven to be more significant than cat states because of their finer phase-space resolution, which distinguishes them in real-world applications. This distinction highlights the growing importance of quantum states capable of achieving enhanced phase-space resolution, offering distinct advantages in various quantum technologies and measurement protocols. The Zurek compass state maintained sensitivity to displacement~\cite{Zurek2001}, which appeared to be associated with the parameter $c_0$; as the amplitude $c_0$ grows, it increases both sensitivity and the average photon number in the state. This suggests that a compass state with a larger average photon number can have greater susceptibility to displacement~\cite{Naeem2021}. Furthermore, in this example, the sensitivity to displacement is anisotropically amplified, as illustrated by the tilelike structures around the origin in Fig.~\ref{fig:fig3} (see Appendix~\ref{appendix:appendixA} for the details). Our main focus is on the case shown in Fig.~\ref{fig:fig1}(a), and as it appears, this particular case is devoid of negative amplitudes, and the dimensions of its phase-space feature are also comparable to those of the coherent states given that sub-Planck structures are absent in this case. Our multiphoton illustrations are specifically devoted to this case, and we will demonstrate how these photon operations will alter its phase-space characteristics. The case depicted in Fig.~\ref{fig:fig1}(a) can be simply called a four-headed kitten state.

\begin{figure}[t]
\includegraphics[width=0.5\textwidth]{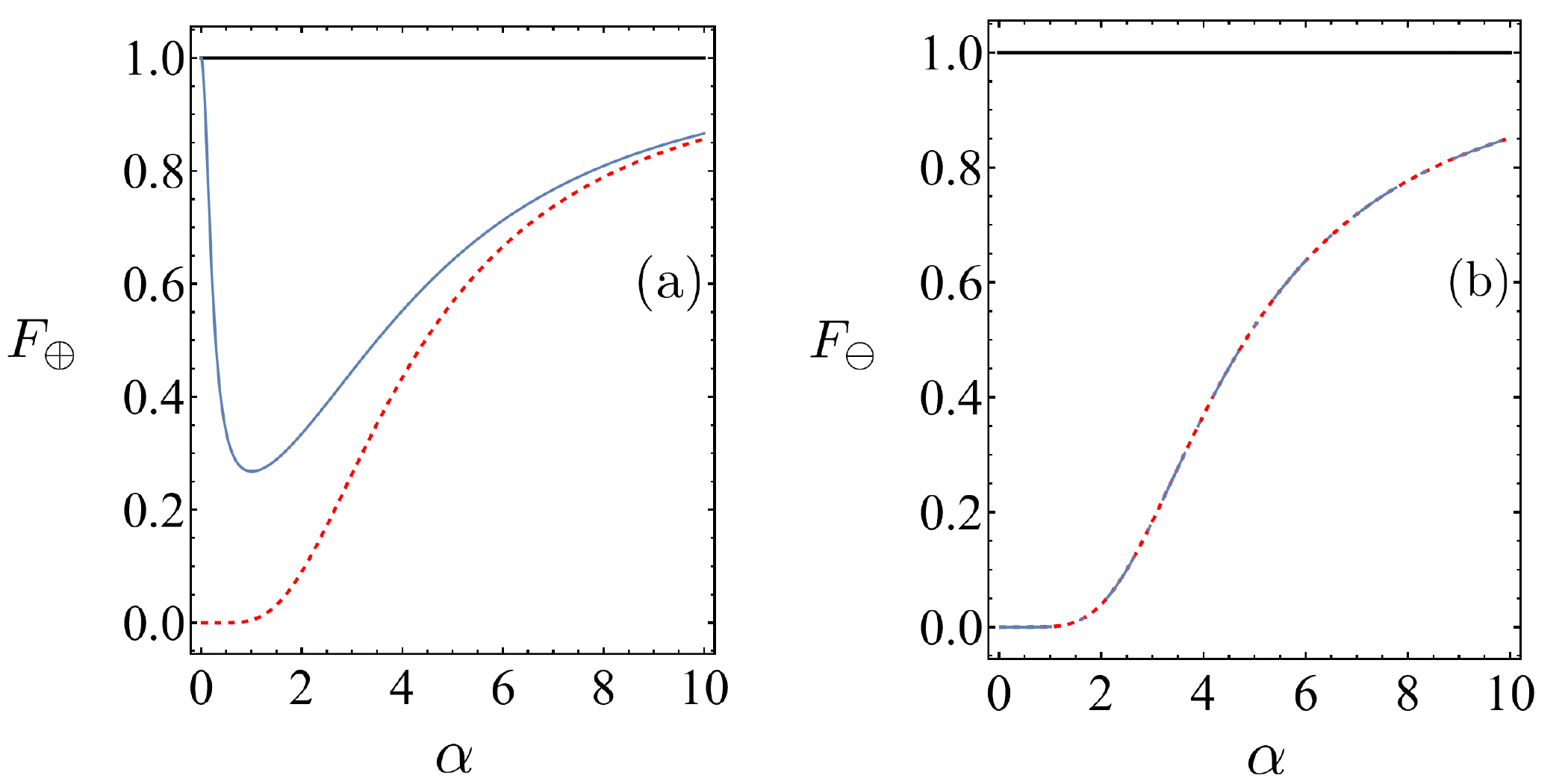}
\caption{Overlap between the coherent and deformed versions. (a) For the SA version of the coherent state, the red dotted line represents $r=4$ and $q=0$ and the blue solid line represents $r=4$ and $q=4$. (b) For the AS version of the coherent state, the red dotted line corresponds to $r=4$ and $q=0$ and the blue dashed line represents $r=4$ and $q=4$. The horizontal black solid lines indicate the case where $r=0$ and $q=0$, representing the overlap between two ordinary coherent states.}
\label{fig:figCohOv}
\end{figure}

\begin{figure*}[t]
\includegraphics[width=\textwidth]{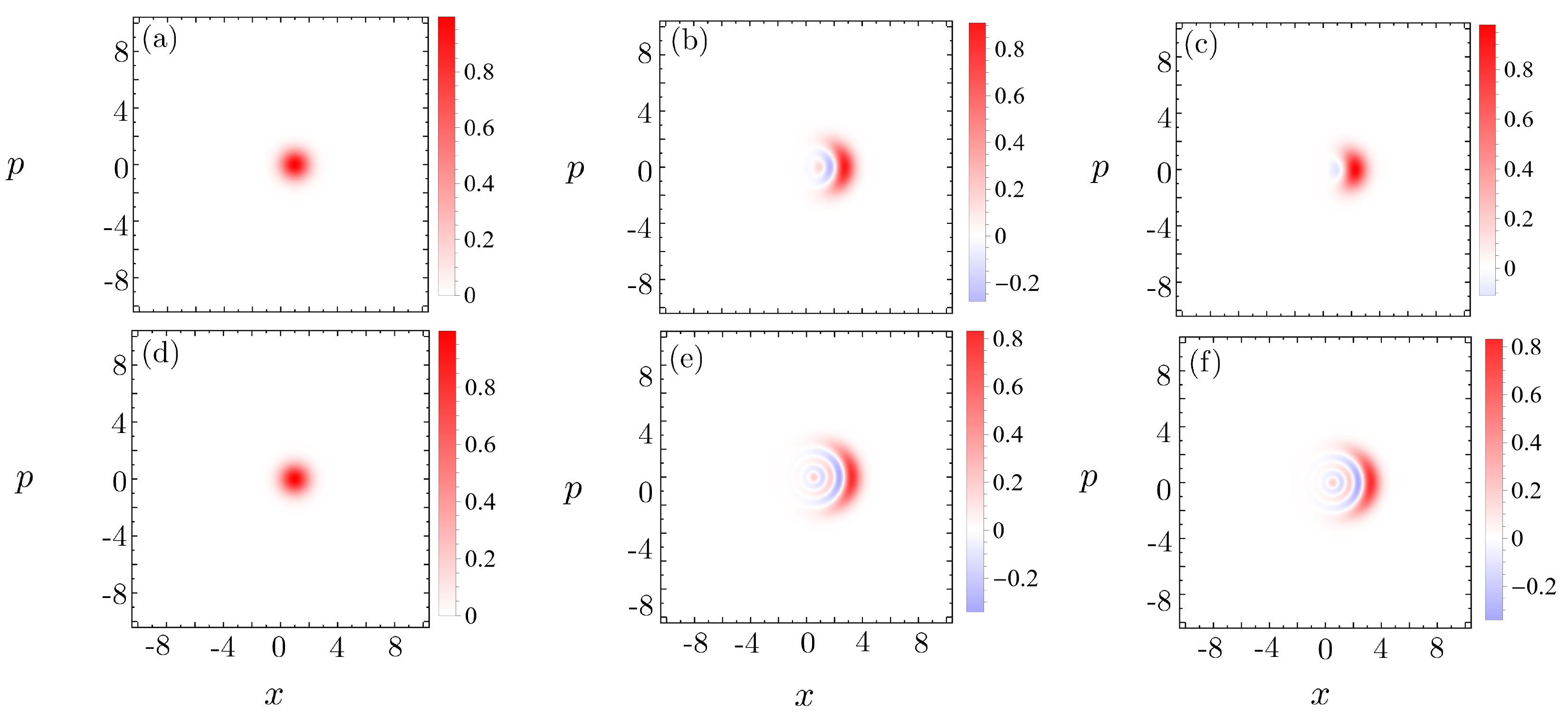}
\caption{(a)-(c) Wigner distributions of the SA case of a coherent state and (d)-(f) corresponding AS cases: (a) $r=0$ and $q=0$, (b) $r=4$ and $q=2$, (c) $r=4$ and $q=4$, (d) $r=0$ and $q=1$, (e) $r=4$ and $q=2$, and (f) $r=4$ and $q=4$. In all cases $\alpha=\nicefrac{1}{\sqrt{2}}$.}
\label{fig:fig4}
\end{figure*}

\section{Refined phase-space densities}\label{sec:photon_varied}

Multiphoton operations have been used both theoritically and experimentally to enhance the phase-space characteristics of a quantum state. For example, techniques involving photon addition (or subtraction) to squeezed-vacuum states have been effectively utilized to generate Schr\"{o}dinger cat states~\cite{Dakna1997, Tang201586, Alexei2006, Neergaard2006} and can also produce multicomponent cat states~\cite{Exp4, Naeem2023}, highlighting the benefits of multiphoton operations. This section focuses on the quantum states obtained by performing photon addition and subtraction operations in different orders.

\begin{figure*}[t]
\includegraphics[width=0.8\textwidth]{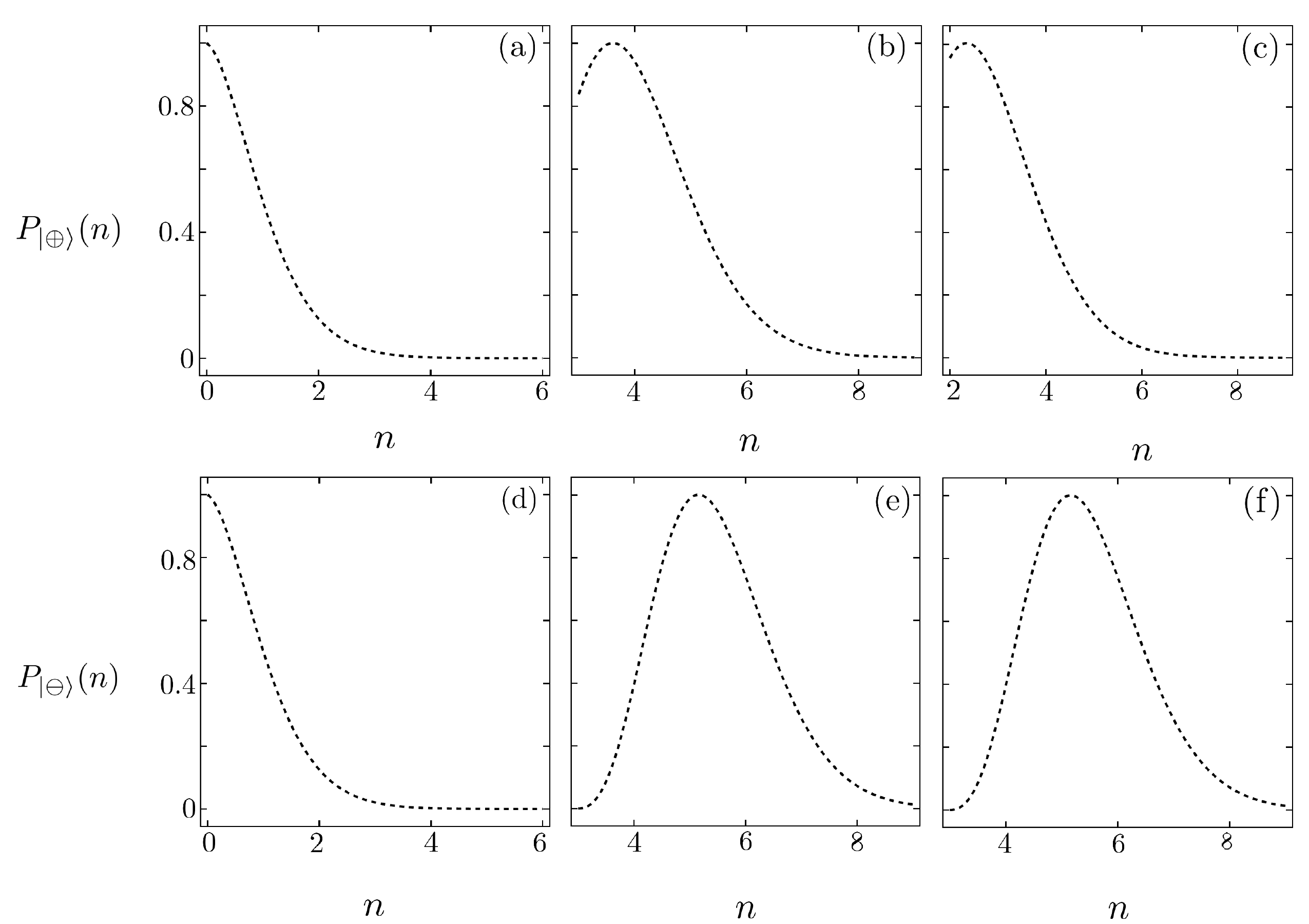}
\caption{PNDs for SA and AS cases of coherent states represented by (a)-(c)~$P_{\ket{\oplus}}(n)$ and (d)-(f)~$P_{\ket{\ominus}}(n)$, respectively, for (a) $r=0$ and $q=0$, (b) $r=4$ and $q=2$, (c) $r=4$ and $q=4$, (d) $r=0$ and $q=2$, (e) $r=4$ and $q=2$, and (f) $r=4$ and $q=4$. We use $\alpha=\nicefrac{1}{\sqrt{2}}$ in all cases, and distributions are normalized to unity.}
\label{fig:fig5}
\end{figure*}

\subsection{Degaussification of coherent states}\label{subsec:d_coh}

In this section, we discuss the deformed coherent states~\cite{Wang11, Barnett2018, Guerrini} obtained by applying a sequence of photon addition and then subtraction, or vice versa, to a standard coherent state. First, consider the case when $r$ photons are added to the coherent state, followed by the subtraction of $q$ photons. This sequence appears as subtraction and addition (SA) operations and mathematically, this case for the coherent $\ket{\alpha}$ is represented as

\begin{align}
    \ket{\oplus}:=N^{-\nicefrac{1}{2}}_{\oplus}\hat{a}^{q}\hat{a}^{\dagger r}\ket{\alpha},
\end{align}
where
\begin{align}
N_{\oplus}=&\nonumber(-1)^{r+q}\sum^{r}_{n=0}  \Gamma H_{r-n,q}\left[\text{i}\alpha,\text{i}\alpha^*\right]\\&\times H_{r-n,q}\left[\text{i}\alpha^*,\text{i}\alpha\right]~\text{with}~\Gamma:=\frac{(-1)^n (r!)^2}{n![(r-n)!]^2}
\end{align}
represents the normalization coefficient and $H_{x,y}$ denotes the bivariate Hermite polynomial.

In the addition-subtraction (AS) scenario, the process involves first subtracting $q$ photons from the state, followed by the addition of $r$ photons, and this case for a coherent state is defined as
\begin{align}
    \ket{\ominus}:=N^{-\nicefrac{1}{2}}_{\ominus}\hat{a}^{\dagger r}\hat{a}^{q}\ket{\alpha},
\end{align}
where
\begin{align}
  N_{\ominus}=|\alpha|^{2q}\sum^r_{n=0}(-1)^n \Gamma |\alpha|^{2(r-n)}
\end{align}
is the corresponding normalization coefficient.

To illustrate the impact of photon addition and subtraction operations on the coherent state $\ket{\alpha}$, we calculate the overlap between the state $\ket{\alpha}$ and its SA and AS variants. The overlap between the coherent state $\ket{\alpha}$ and its SA version $\ket{\oplus}$ is defined as $F_{\oplus}(\alpha) := |\braket{\alpha|\oplus}|^2$, while the overlap in the AS case is defined as $F_{\ominus}(\alpha) :=| \braket{\alpha|\ominus}|^2$. These overlaps are evaluated numerically and are represented in Fig.~\ref{fig:figCohOv}, where it is evident that both SA and AS situations differ significantly from the coherent state in lower $\alpha$ ranges, and as $\alpha$ grows, corresponding SA and AS variants of the coherent state simply transform back to the ordinary coherent state. This comparison is carried out with different amounts of photon addition $r$ and subtraction $q$, as shown in Fig.~\ref{fig:figCohOv}(a) and Fig.~\ref{fig:figCohOv}(b), respectively. Note that the photon subtraction operation in the AS case has no effect, and it essentially represents the photon-added version of the coherent state. This is justified in Fig.~\ref{fig:figCohOv}(b), where the red dotted line and blue dashed line correspond to different $q$ but same $r$ values, resulting in the same curve.

The Wigner functions of SA and AS cases are evaluated using Eq.~(\ref{eq:Wig1}). Compared to their counterparts of ordinary coherent states, these Wigner functions attain non-Gaussian form and may exhibit negative phase-space attributes for different values of $r$ and $q$, highlighting their nonclassical nature~\cite{Wang11, PhysRevA.102.032413, Chabaud2021}.
First, for the SA example, the corresponding Wigner function is derived as
\begin{align}\label{eq:pa_wig_coh}
 W_{\ket{\oplus}}(\beta)=&\nonumber \frac{1}{N_{\oplus}} W_{\ket{\alpha}\bra{\alpha}}(\beta)\sum^{r}_{n=0} \Gamma  H_{r-n,q}\left[-\text{i}\Omega,-\text{i}\alpha^*\right]\\&\times H_{r-n,q}\left[\text{i}\Omega^*,\text{i}\alpha\right],
\end{align}
where $\Omega=\nonumber 2\beta-\alpha$. The non-Gaussian nature of this Wigner function is obvious from the expression (\ref{eq:pa_wig_coh}), and it is also evident in Figs.~\ref{fig:fig4}(a)-\ref{fig:fig4}(c) that this Wigner function has now acquired negative amplitudes, reflecting the nonclassical nature of the SA case of the coherent state. Note that Fig.~\ref{fig:fig4}(a) with $r=0$ and $q=0$, represents the case of the corresponding ordinary coherent state, but as observed in Fig.~\ref{fig:fig4}(b) for nonzero $r$ and $q$ values, the corresponding Wigner function holds negative amplitudes, and then in Fig.~\ref{fig:fig4}(c) an increment in $q$ values with a constant $r$ reduces the negative amplitudes in the Wigner function. For this case, it can be shown that higher $r$ amplifies negative regions while larger $q$ may cause the elimination of the negative amplitudes in the Wigner phase space.

Mathematically, the Wigner function for the AS case has a form analogous to that of the SA case, that is,
\begin{align}
   W_{\ket{\ominus}}(\beta)= &\nonumber\frac{(-1)^r}{N_{\ominus}} W_{\ket{\alpha}\bra{\alpha}}(\beta)\sum^r_{n=0}(-1)^n \Gamma(2\beta-\alpha)^{r-n}\\&\times(\alpha^*-2\beta^*)^{r-n}.
\end{align}
This Wigner function is shown in Figs.~\ref{fig:fig4}(d)-\ref{fig:fig4}(f) with different $r$ and $q$ values, indicating that the corresponding Wigner functions also contain negative amplitudes in the phase space. Note that in this case $q$ has no effect on the state, as shown in Fig.~(\ref{fig:fig4})(d), with $r=0$ and $q=1$ appearing to have the same Wigner function as a coherent state. This is additional evidence that the photon subtraction effect in the indicated AS case is zero, and we know that it simply reflects the photon-added case. The photon subtraction from a coherent state leaves the state unchanged, as has been proved experimentally~\cite{Zavatta_2008}, as the coherent state is an eigenstate of the annihilation operator. As illustrated in Fig.~(\ref{fig:fig4})(e), similar to the SA case, the photon addition in this case also enhances the negative regions. Furthermore, as illustrated in Fig.~(\ref{fig:fig4})(f), where the parameter $r$ is kept the same as in the previous case and $q$ is increased, again it appears that the increase in $q$ (the number of photon subtraction) in the AS case has no effect on the Wigner distribution.

In summary, both SA and AS cases of the coherent state are non-Gaussian, and the negativity in their Wigner functions confirms the nonclassical nature of these states, which is lacking in the original coherent states. The addition and subtraction (or subtraction and addition) of an equal number of photons from a quantum state can result in two different quantum states. This is confirmed here by two AS and SA cases of coherent states with equivalent photon operations resulting in distinct quantum states, as shown by their Wigner function illustrations, and can also be confirmed by the noncommutativity of the bosonic operators $\hat{a}^{\dagger}$ and $\hat{a}$.

\begin{figure}[t]
\includegraphics[width=0.5\textwidth]{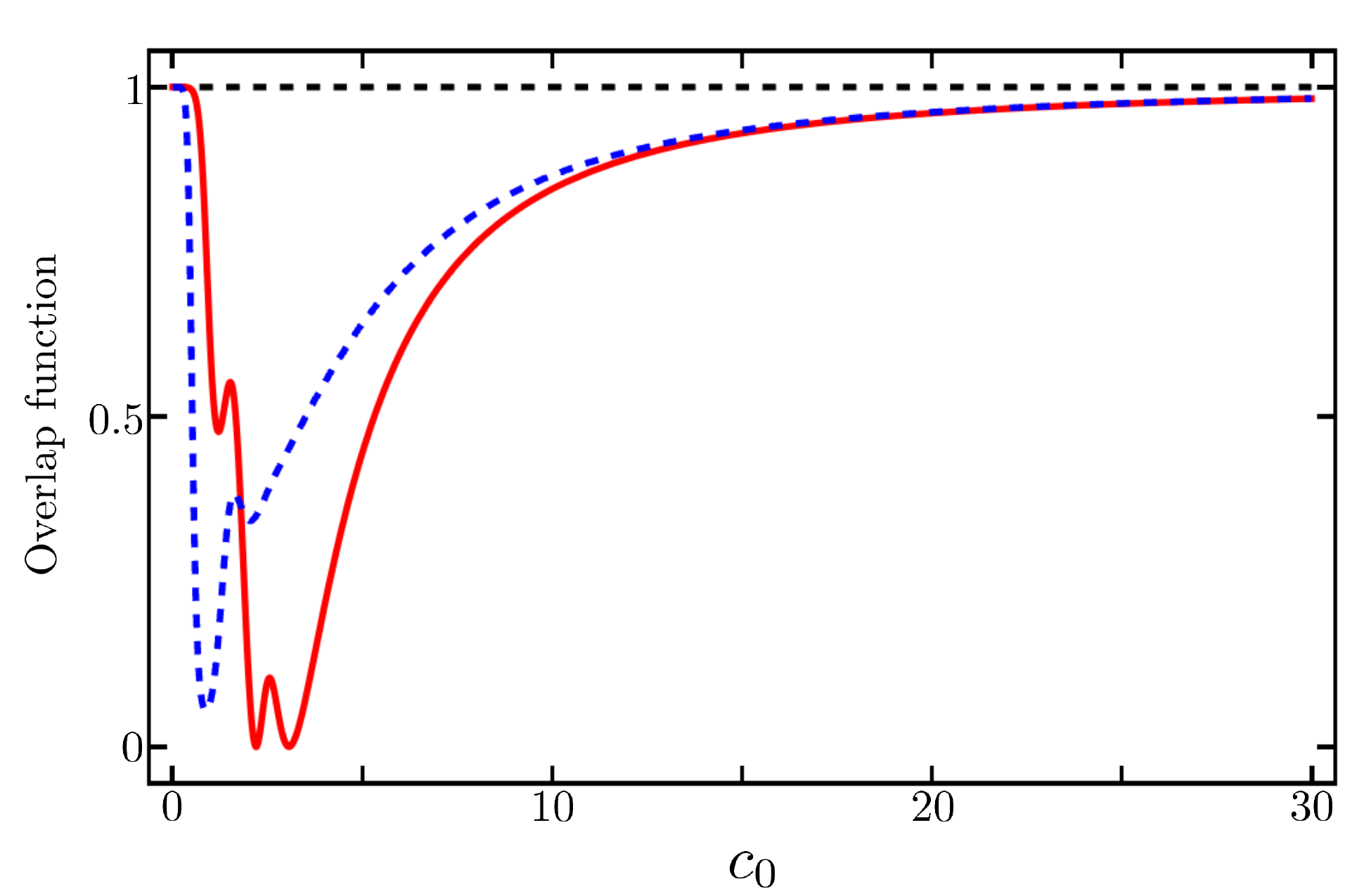}
\caption{Overlap function $|F_{\square}(c_0)|^2$ (blue dashed curve) and $|F_{\blacksquare}(c_0)|^2$ (red solid line), with $r=4$ and $q=4$. The horizontal black dashed line depicts the case where $r=0$ and $q=0$; consequently, the overlap is simply between two compass states.}
\label{fig:fig6}
\end{figure}

\subsection{Fluctuations in photon numbers}\label{PND}

In this section, we explore how the order in which photon addition and subtraction operations are applied to a state influences the photon number statistics. The photon addition and subtraction operations, as examined in our case, are anticipated to affect the photon number distribution (PND) of our quantum states. For example, in the compass state, the sizes of sub-Planck structures, phase-space sensitivity, and the PND are all proportional to the macroscopic parameter $c_0$~\cite{Naeem2021}. We now demonstrate the consequences of photon operations with varying values $r$ and $q$, and for this we evaluate the corresponding PND of each case of reduced coherent states. To investigate the PND in these SA and AS scenarios of the coherent state, we use the mathematical expression for the PND of a quantum state $\ket{\psi}$, defined as $P_{\ket{\psi}}(n):=\left|\braket{n|\psi}\right|^2$, where $\ket{n}$ represents the Fock state~\cite{Sch01}. 

The PND for the SA and AS cases, denoted as $P_{\ket{\oplus}}(n)$ and $P_{\ket{\ominus}}(n)$, are evaluated as
\begin{align}
    P_{\ket{\oplus}}(n)=\frac{N_\oplus[(q+n)!]^2 \kappa}{n![(q+n-r)!]^2}
\end{align}
and
\begin{equation}
    P_{\ket{\ominus}}(n)=\frac{N_\ominus n! \kappa}{[(n-r)!]^2}
\end{equation}
with
\begin{align}
    \kappa:=|\alpha|^{2(q-r+n)}\text{e}^{-|\alpha|^2}.
\end{align}
Let us now analyze these distributions. Figures~\ref{fig:fig5}(a)-\ref{fig:fig5}(c) show the PNDs for the SA case of the coherent state for different situations based on the varying amount of added and subtracted photons. As depicted in Fig.~\ref{fig:fig5}(a), the PND with $r=0$ and $q=0$ exhibits a Poissonian distribution, which obviously corresponds to a coherent state. In Fig.~\ref{fig:fig5}(b), when $r=4$ photons are added and we set $q=0$, the Poissonian distribution shifts to higher values of $n$, with the peak now occurring at a larger mean photon number. In Fig.~\ref{fig:fig5}(c), the Poissonian distribution shifts to lower values of $n$, with an incremental subtraction of photons to $q=4$, and the number of added photons is kept at $r=4$ as in the prior instance. These cases reflects that in the SA examples, a higher $r$ corresponds to a higher average photon number, whereas a higher $q$ corresponds to a lower average photon number in the resultant state. 

\begin{figure*}
\includegraphics[width=1.01\textwidth]{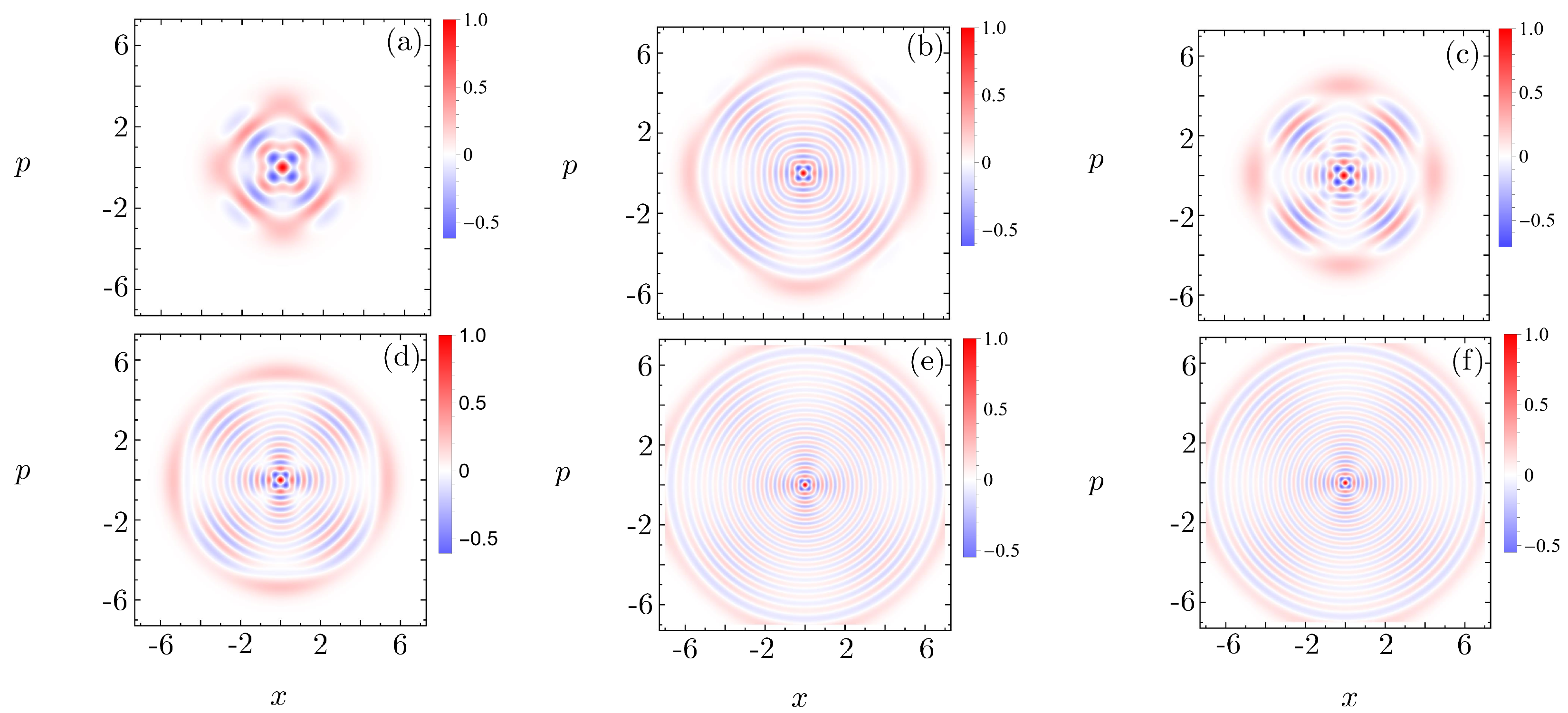}
\caption{[(a)-(c)] the Wigner function of the SA case of the four-headed kitten state and [(d)-(f)] the analogous AS cases for (a)~$r=12$ and $q=12$, (b)~$r=24$ and $q=12$, (c)~$r=24$ and $q=20$, (d)~$r=12$ and $q=12$, (e)~$r=24$ and $q=12$, and (f)~$r=24$ and $q=20$. In all cases, $c_0 = 1$.}
\label{fig:fig8}
\end{figure*}

\begin{figure}[t]
\includegraphics[width=0.47\textwidth]{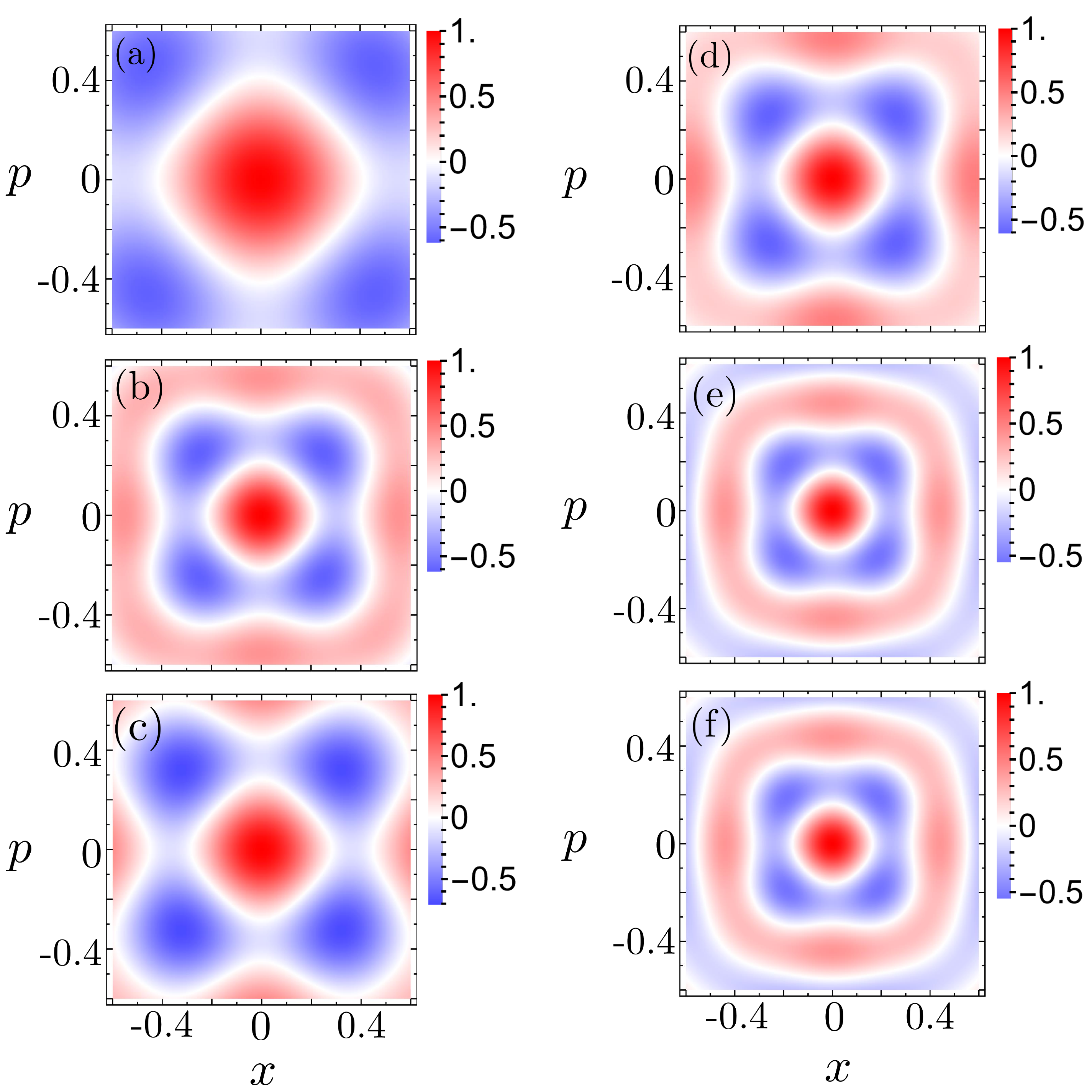}
\caption{Close-up of the central phase-space structures of the cases shown in Fig.~\ref{fig:fig8}. [(a)-(c)] SA and [(d)-(f)] AS for (a) $r=12$ and $q=12$, (b) $r=24$ and $q=12$; (c) $r=24$ and $q=20$, (d) $r=12$ and $q=12$, (e) $r=24$ and $q=12$ (f) $r=24$ and $q=20$. In each example, $c_0 = 1$.}
\label{fig:fig9}
\end{figure}

The PND for the AS case is shown in Figs.~\ref{fig:fig5}(d)-\ref{fig:fig5}(f) for a few $r$ and $q$ values. As illustrated in Fig.~\ref{fig:fig5}(d), for $q=2$ and $r=0$, the statistics of the PND stay invariant, showing that raising the number of subtracted photons has no influence on the average photon number of the states when applied directly to a coherent state. As observed in Fig.~\ref{fig:fig5}(e), increasing the number of added photons while maintaining the number of subtracted photons constant, that is, with $r=4$ and $q=2$, drives the Poissonian distribution to a larger $n$. In Fig.~\ref{fig:fig5}(f), the number of subtracted photons is increased to $q=4$ without changing $r$, and it is observed that subtracting photons from a coherent state has no influence on the related PND. This is an additional confirmation of how photon subtraction operations keep the PND of this case invariant. Hence, photon addition in both the SA and AS situations increases the average photon number in the coherent states; however, the AS case maintains the distribution at higher mean photon values, implying that this situation has higher average photon numbers than the SA case of the coherent state.

\subsection{Impact of multiphoton operations}

In \S\ref{subsec:d_coh}, we noticed that the SA and AS variants of a standard coherent state exhibit nonclassical phase-space features, and it is observed that the sequence in which photon operations are applied significantly affects the outcome, resulting in two distinct quantum states with different phase-space characteristics. We now extended those basic notions to our main quantum states of the present work.
First, for our SA scenario, $r$ photons are added to the compass state $\ket{\diamondsuit}$ as given in Eq.~(\ref{eq:compass0_state}), followed by the subtraction of $q$ photons. Conversely, in the AS scenario, $q$ photons are first subtracted from the state $\ket{\diamondsuit}$, and then $r$ photons are added.

The mathematical representation of the SA case of the state $\ket{\diamondsuit}$ is
\begin{equation}\label{eq:SA_kitten}
    \ket{\square}:=N_{\square}^{-\nicefrac{1}{2}}\hat{a}^{q}\hat{a}^{\dagger r}\sum^{4}_{i=1}\ket{\alpha_i}
\end{equation}
with
\begin{align}
    N_{\square}=&\nonumber (-1)^{r+q}\sum^4_{i,j=1}\sum^{r}_{n=0}\Gamma G_{\alpha_i,\alpha_j}\text{e}^{\alpha^*_i\alpha_j} H_{r-n,q}\left[\text{i}\alpha_j,\text{i}\alpha^*_i\right]\\&\times H_{r-n,q}\left[\text{i}\alpha^*_i,\text{i}\alpha_j\right]
\end{align}
the normalization coefficient.

In the same way, the AS case is obtained as
\begin{align}\label{eq:AS_kitten}
    \ket{\blacksquare}:=N_{\blacksquare}^{-\nicefrac{1}{2}}\hat{a}^{\dagger r}\hat{a}^{q}\sum^{4}_{i=1}\ket{\alpha_i},
\end{align}
where
\begin{align}
   N_{\blacksquare}=&\nonumber \sum^4_{i,j=1}\sum^{r}_{n=0}(-1)^n \Gamma G_{\alpha_i,\alpha_j}\text{e}^{\alpha^*_i\alpha_j}(\alpha^*_i \alpha_j)^q \\&\times(\alpha_j)^{r-n}(\alpha^*_i)^{r-n}
\end{align}
represents the normalization factor for this case.

We now compare the original state $\ket{\diamondsuit}$ with the proposed variants by assessing their overlap. This overlap measures distinctness between  the proposed states and the original compass state. The overlap between $\ket{\square}$ and $\ket{\diamondsuit}$ is defined as $F_{\square}(c_0):=\braket{\square|\diamondsuit}$. This overlap reads
\begin{align}
F_{\square}(c_0)=(-\text{i})^{r+q}\left(N_{\square}N_{\diamondsuit}\right)^{\nicefrac{-1}{2}}\sum^4_{i,j=1}F_{\ket{\square_i}\bra{\square_j}}
\end{align}
with
\begin{align}
F_{\ket{\square_i}\bra{\square_j}}:=G_{\alpha_i,\alpha_j}\text{e}^{\alpha^*_i\alpha_j}H_{r,q}\left[\text{i}\alpha^*_i,\text{i}\alpha_j\right].
\end{align}
Similarly, for the AS case, the overlap function $F_{\blacksquare}(c_0):=\braket{\blacksquare|\diamondsuit}$ is
\begin{align}
F_{\blacksquare}(c_0)=\left(N_{\blacksquare}N_{\diamondsuit}\right)^{\nicefrac{-1}{2}}\sum^4_{i,j=1}F_{\ket{\blacksquare_i}\bra{\blacksquare_j}},
\end{align}
where
\begin{align}
F_{\ket{\blacksquare_i}\bra{\blacksquare_j}}:=G_{\alpha_i,\alpha_j}\text{e}^{\alpha^*_i\alpha_j}(\alpha^*_i)^q(\alpha_j)^r.
\end{align}
The corresponding overlaps are shown in Fig.~\ref{fig:fig6} for $r=q=4$, where the red solid line represents the AS case and the blue dashed line represents the overlap SA case. The black dashed line represents the case when $r=q=0$ illustrating the case of the overlap between two same compass states $\ket{\diamondsuit}$. For smaller values of $c_0$, the overlaps between the states show a more significant difference, highlighting greater distinctions between the comparable states. However, for larger $c_0$ values, the blue dashed and red solid lines converge to the same level as the horizontal black dashed line, indicating that both the SA and AS cases return to the original compass state.

In our work, we stick with the lower $c_0$ situation, which essentially belongs to the case of the four-headed kitten state, and then we investigate the corresponding phase-space characteristics of the resultant states obtained by applying a large amount of photon addition (and subtraction) to the indicated four-headed kitten state. Here, note that the AS case of the four-headed kitten state is the eigenstate of the operator $\hat{a}^4$, i.e., $\hat{a}^4 \ket{\blacksquare} = \ket{\diamondsuit}$. This shows that choosing the photon subtraction $q$ as a multiple of 4 ($q=4k$ with $k\in \mathbb{N}^+$) corresponds to a photon-added version of the four-headed kitten state. When $p = 0$, this reduces to the ordinary case of the four-headed kitten state denoted by $\ket{\diamondsuit}$.

\subsection{Unique phase-space features}\label{subsec:Wig_var}

In the preceding section we introduced the SA and AS cases of our interest, each producing distinct quantum states as observed by their overlap, and now in this section, we particularly employ the Wigner function to investigate the corresponding phase space of these quantum states. To obtain the Wigner function for each case, Eq.~(\ref{eq:Wig1}) is employed, and we denote by $W_{\ket{\square}}(\beta)$ and $W_{\ket{\blacksquare}}(\beta)$ as the corresponding Wigner functions of the SA and AS cases of the kitten state, respectively.

\begin{figure}
\includegraphics[width=0.47\textwidth]{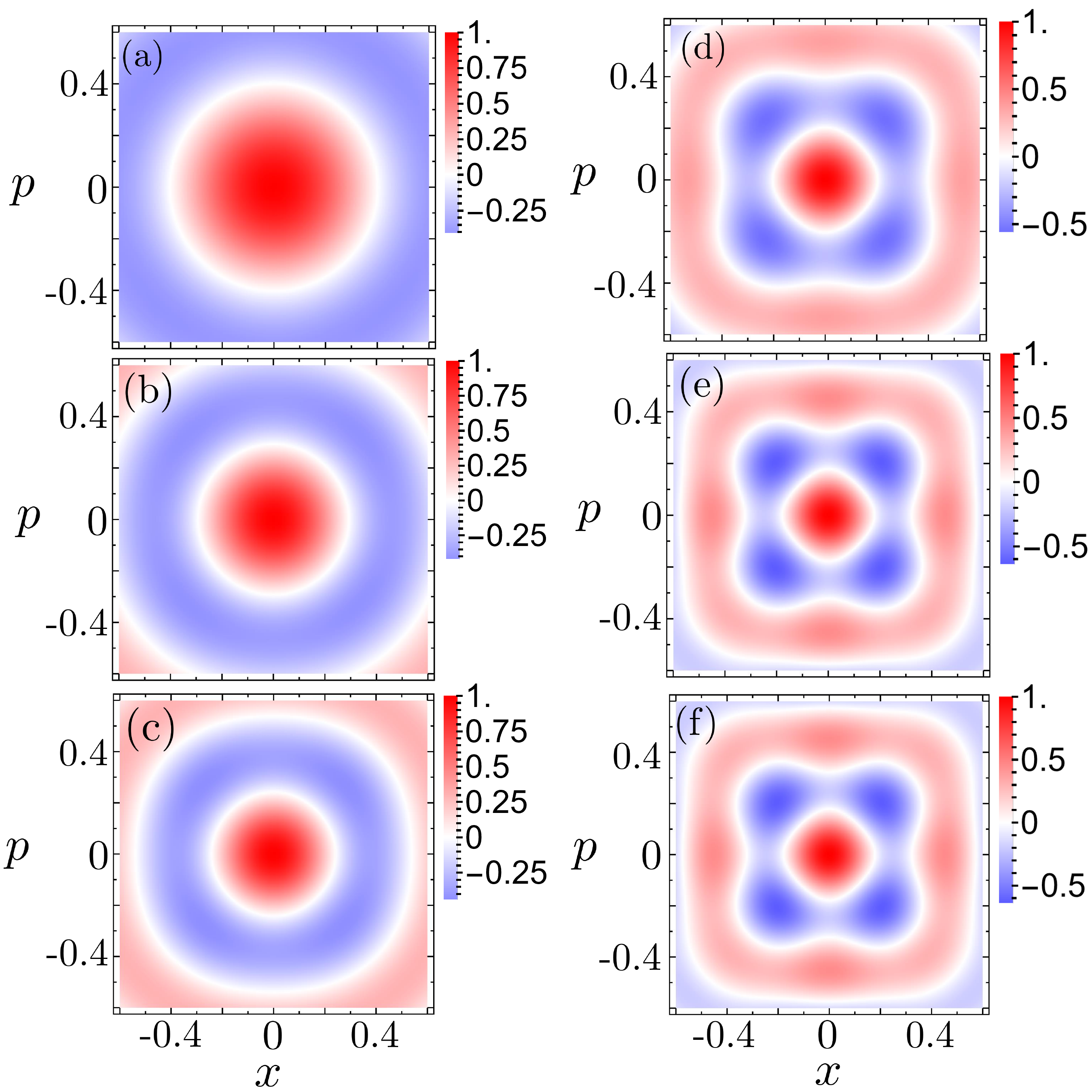}
\caption{Close-up of the center phase-space structures of the AS case of the four-headed kitten state with multiple values of the parameters $r$ and $q$: (a) $r=1$ and $q=1$, (b) $r=5$ and $q=5$, (c) $r=9$ and $q=9$, (d) $r=16$ and $q=10$, (e) $r=22$ and $q=10$, (f) $r=22$ and $q=18$. For all cases, $c_0 = 1$.}
\label{fig:fig10}
\end{figure}

Let us now examine the Wigner distributions for each scenario. For the SA case of our four-headed kitten state introduced in Eq.~(\ref{eq:SA_kitten}), the Wigner distribution is calculated as
\begin{align}
    W_{\ket{\square}}(\beta)=\frac{1}{N_{\square}}\sum^4_{i,j=1} W_{\ket{\square_i}\bra{\square_j}}(\beta),
\end{align}
where
\begin{align}\label{eq:pa_wig}
 W_{\ket{\square_i}\bra{\square_j}}(\beta):=&\nonumber W_{\ket{\alpha_i}\bra{\alpha_j}}(\beta)\sum^{r}_{n=0} \Gamma H_{r-n,q}\left[\text{i}\Omega^*_j,\text{i}\alpha_i\right]\\&\times H_{r-n,q}\left[-\text{i}\Omega_i,-\text{i}\alpha^*_j\right]
\end{align}
with
\begin{align}
    \Omega_\mu:=2\beta-\alpha_\mu.
\end{align}
In the same way, for the AS situation depicted in Eq.~(\ref{eq:AS_kitten}), we have
\begin{align}
W_{\ket{\blacksquare}}(\beta)=\frac{(-1)^r}{N_{\blacksquare}}\sum^4_{i,j=1} W_{\ket{\blacksquare_i}\bra{\blacksquare_j}}(\beta),
\end{align}
where
\begin{align}\label{eq:ps_wig}
 W_{\ket{\blacksquare_i}\bra{\blacksquare_j}}(\beta):=&\nonumber (\alpha_i \alpha^*_j)^q W_{\ket{\alpha_i}\bra{\alpha_j}}(\beta)\sum^{r}_{n=0} (-1)^n\Gamma (2\beta-\alpha_i)^{r-n}\\&\times (\alpha^*_j-2\beta^*)^{r-n}.
\end{align}
The corresponding Wigner functions are shown in Fig.~\ref{fig:fig8}, with Fig.~\ref{fig:fig9} illustrating a close-up of the central phase-space features of each case; Figs.~\ref{fig:fig8}(a)-\ref{fig:fig8}(c) and \ref{fig:fig9}(a)-\ref{fig:fig9}(c) exhibit the SA, while Figs.~\ref{fig:fig8}(d)-\ref{fig:fig8}(f) and \ref{fig:fig9}(d)-\ref{fig:fig9}(f) provide equivalent AS cases of the four-headed kitten state. With specific selections of the parameters $r$ and $q$, Fig.~\ref{fig:fig10} offers a further illustration of the Wigner function of corresponding AS scenarios. 
It is readily apparent that our SA and AS instances achieve substantially distinct phase-space characteristics as compared to the four-headed kitten state. Interestingly, a central sub-Planck structure is identified in each of the cases outlined. We focus on the significance of the sub-Planck structure in these states and examine this particular phase-space feature in detail.

Photon addition and subtraction appear to have significant effects on the Wigner distribution of the corresponding states. Note that the original four-headed kitten state, represented in Fig.~\ref{fig:fig1}(a) with $c_0=1$, does not exhibit a sub-Planck structure. This implies that the appearance of the sub-Planck structures in our cases is attributed to the photon operations involved. Specifically, it is observed that as the parameter $r$, which represents the number of added photons, increases, the dimensions of the sub-Planck structure reduce uniformly in both the SA and AS cases and hence can be far smaller than the extension found for the coherent state when $r\gg1$. This is evident by comparing the scenarios presented in Figs.~\ref{fig:fig9}(a) and \ref{fig:fig9}(b) for the SA case, and then Figs.~\ref{fig:fig9}(d) and \ref{fig:fig9}(e) are for the corresponding AS cases, where increasing $r$ with constant $q$ clearly depicts this impact. Furthermore, increasing the number of photon subtractions $q$ increases the size of the sub-Planck structure in the SA case, as observed by comparing cases depicted in Figs.~\ref{fig:fig8}(b) and \ref{fig:fig8}(c), while it has no effect in the AS case, as shown in Figs.~\ref{fig:fig8}(e) and \ref{fig:fig8}(f). Note that all the examples of the AS described in Figs.~\ref{fig:fig8} and \ref{fig:fig9} correspond to the merely photon-added case of the compass state, as for these situations $q$ is a multiple of 4, where the corresponding four-headed kitten state is the eigenstate of the operator $\hat{a}^4$.

We find that photon addition generally reduces the size of sub-Planck structures in both SA and AS scenarios, while photon subtraction increases their size in the SA case but has no effect on their size in the AS case. To further illustrate this for the AS case, we examine situations where the number of photon subtractions is not a multiple of 4, as shown in Fig.~\ref{fig:fig10}. Increasing the photon addition to the state as previously noted also reduces the size of the sub-Planck structures for this specific illustration [Figs.~\ref{fig:fig10}(d) and \ref{fig:fig10}(e)] while as expected, varying the photon subtraction in these cases has no impact on the size of the sub-Planck structures [Figs.~\ref{fig:fig10}(e) and \ref{fig:fig10}(f)], but in some specific instances, photon subtraction slightly enhances the isotropy of the sub-Planck structures. For example, an enhancement in the isotropy of the central sub-Planck structures of these cases is clearly evident in Figs.~\ref{fig:fig10}(a) to \ref{fig:fig10}(c), as demonstrated by the circular structure around the origin, which is further analyzed in the Appendix~\ref{appendix:appendixA}. This indicates that, for these scenarios, the sub-Planck structure is uniformly constrained in all directions of the phase space, hence representing an isotropic version of the sub-Planck structures (see the Appendix~\ref{appendix:appendixA}).

\section{Enhancement in sensitivity}\label{subsec:sens_var}

Sensitivity and its relationship to the phase-space characteristics of a quantum state were thoroughly discussed in the \S\ref{subsec:main_theory}. These concepts were then applied to the compass state in \S\ref{subsec:compass}, where it was demonstrated that the presence of sub-Planck structures in those states significantly enhances the sensitivity to displacement and that varying controlling parameters may further enhance this sensitivity far better than the standard quantum limit.
In \S\ref{subsec:Wig_var}, we thoroughly examined the phase space of the proposed SA and AS cases of the four-headed kitten state, confirming the presence of sub-Planck structures in their phase spaces. We now examine how these sub-Planck structures have an impact on the sensitivity to phase-space displacement, which is analyzed by assessing the sensitivities using Eq.~(\ref{eq:sens1}).

\begin{figure}
\includegraphics[width=0.5\textwidth]{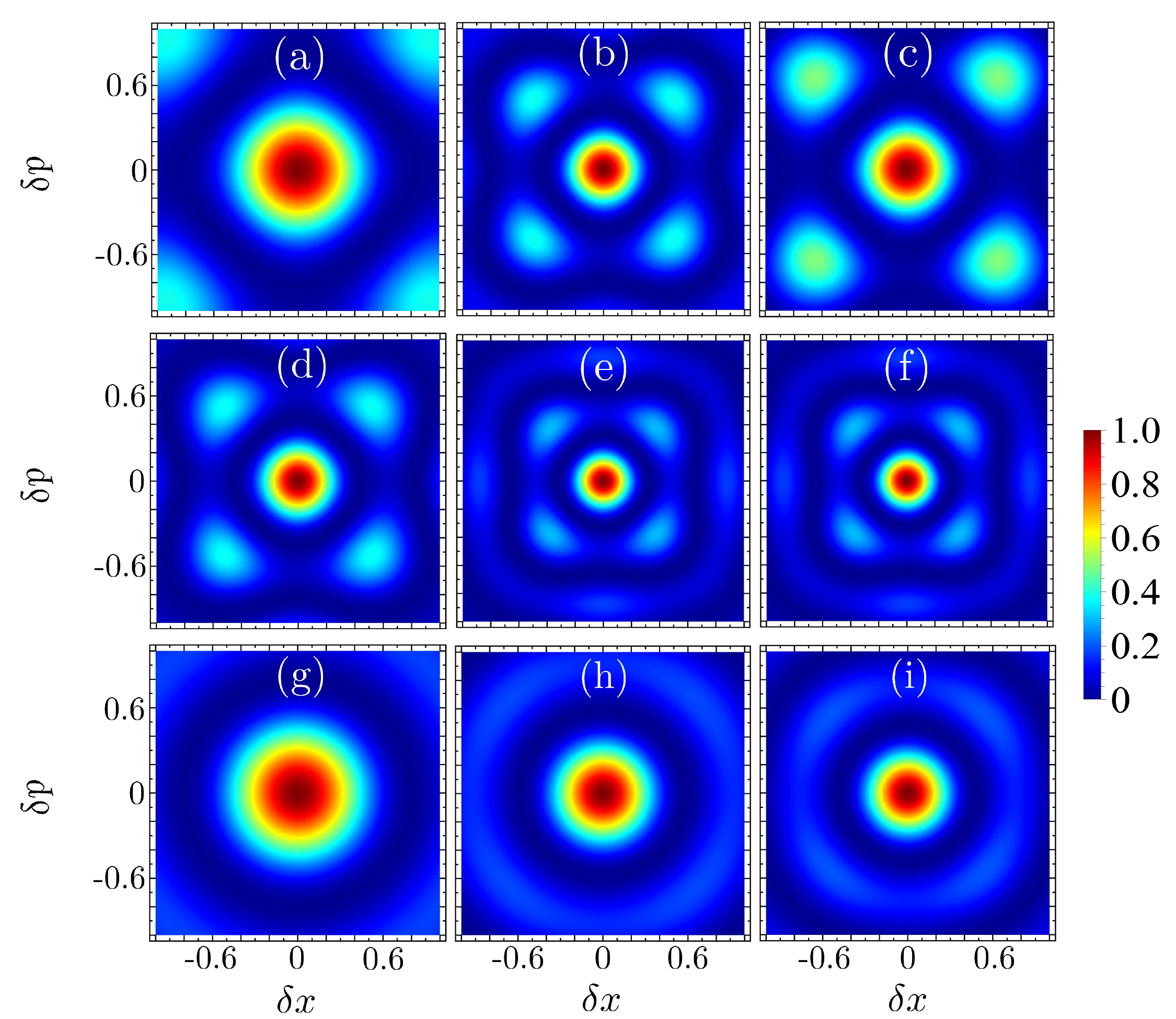}
\caption{Overlap between a state and its slightly translated version for [(a)-(c)] SA examples and [(d)-(i)] comparable AS cases: (a)~$r=12$ and $q=12$, (b)~$r=24$ and $q=12$, (c)~$r=24$ and $q=20$, (d)~$r=12$ and $q=12$, (e)~$r=24$ and $q=12$, (f)~$r=24$ and $q=20$, (g)~$r=q=1$, (h)~$r=q=5$, and (i)~$r=q=9$. In all situations, we set $c_0 = 1$ and normalize the intensity plots to unity.}
\label{fig:fig11}
\end{figure}

First, for the SA scenario, we denote the associated sensitivity by $S_{\ket{\square}}(\delta)$, which is calculated as
\begin{align}
S_{\ket{\square}}(\delta)=\left|\sum^4_{i,j=1}O_{\ket{\square_i}\bra{\square_j}}(\delta)\right|^2
\end{align}
with
\begin{align}
O_{\ket{\square_i}\bra{\square_j}}(\delta):=&\nonumber G_{\alpha_i,\alpha_j}\Lambda \sum^{r}_{n=0} (-1)^n\Gamma H_{r-n,q}\left[\text{i}\big(\alpha^*_i-\delta^*\big),\text{i}\alpha_j\right]\\&\times H_{r-n,q}\left[-\text{i}\big(\alpha_j+\delta\big),-\text{i}\alpha^*_i\right]
\end{align}
and
\begin{align}
 \Lambda:=\exp\left[-\alpha_j \delta^*-\frac{|\delta|^2}{2}+\alpha^*_i\alpha_j+\alpha^*_i\delta\right].
\end{align}
In the AS situation, the sensitivity $S_{\ket{\blacksquare}}(\delta)$ is found as
\begin{align}
S_{\ket{\blacksquare}}(\delta)=\left|\sum^4_{i,j=1}O_{\ket{\blacksquare_i}\bra{\blacksquare_j}}(\delta)\right|^2,
\end{align}
where
\begin{align}
 O_{\ket{\blacksquare_i}\bra{\blacksquare_j}}(\delta):=&\nonumber (\alpha^*_i \alpha_j)^qG_{\alpha_i,\alpha_j}\Lambda \sum^{r}_{n=0}(-1)^n \Gamma  (\alpha^*_i-\delta^*)^{r-n}\\&\times(\alpha_j+\delta)^{r-n}.
\end{align}
The corresponding sensitivities $S_{\ket{\square}}(\delta)$ and $S_{\ket{\blacksquare}}(\delta)$ with $\delta:=\nicefrac{(\delta x+\text{i}\delta p)}{\sqrt{2}}$ are shown in Fig.~\ref{fig:fig11}.

Let us now examine the SA cases presented in Figs.~\ref{fig:fig11}(a)-\ref{fig:fig11}(c), where it is evident that the overlap is zero for values $|\delta|<1$ (less than a coherent state) along arbitrary directions in phase space. This concept is also illustrated in detail in Appendix \ref{appendix:appendixA}, which reflects that the sensitivity to displacement in this scenario surpasses the standard limit. The enhancement in sensitivity becomes more pronounced as the parameter $r$, which represents the number of added photons to the four-headed kitten state, increases. As shown in Figs.~\ref{fig:fig11}(a) and \ref{fig:fig11}(b), this effect becomes clearly noticeable with a higher number of photons added. Specifically, the central structure is significantly reduced when the number of added photons increases from $r=12$ to $r=24$ while the number of subtracted photons $q=12$ remains constant in this situation. Furthermore, the impact of increasing the number of eliminated photons from $q=12$ to $q=20$ is illustrated in the Figs.~\ref{fig:fig11}(b) and \ref{fig:fig11}(c), where an enlargement in the central structure occurs, indicating that a higher value of $|\delta|$ relative to the earlier case represented in Fig.~\ref{fig:fig11}(b) is needed to make the overlap zero. This suggests that the sensitivity to displacement in this circumstance decreases with an increment in $q$; hence, contrary to the parameter $r$, increasing the number of subtracted photons reduces sensitivity in the SA scenario.

In the AS scenario, as depicted in Figs.~\ref{fig:fig11}(d)-\ref{fig:fig11}(f), comparable to the SA example, an improved sensitivity to phase space displacements $\delta$ is observed. The values $|\delta|<1$ can render the $S_{\ket{\blacksquare}}(\delta)$ zero as observable by the central structure. Similar to the SA example, this amplification becomes more noticeable as $r$ increases, as illustrated in the instances depicted in Figs.~\ref{fig:fig11}(d) and \ref{fig:fig11}(e), where an increment in $r$ as shown in Fig.~\ref{fig:fig11}(e) has reduced the central structure corresponding to the overlap function, meaning that now a smaller value of $|\delta|$ is required to make the overlap zero compared to the case shown in Fig.~\ref{fig:fig11}(d).
Furthermore, we observe that in Figs.~\ref{fig:fig11}(e) and \ref{fig:fig11}(f), the photon subtraction $q$ does not reduce the sensitivity enhancement, increasing $q$ results in the same overlap as its preceding instance. However, the photon subtraction $q$ may have an effect on the isotropic nature of the sensitivity, as optimal parameter selections may produce isotropic regions. For example, the circle-type regions centered at the origin observed in Figs.~\ref{fig:fig11}(g)-\ref{fig:fig11}(i) show that in the AS scenario, sensitivity is consistently increased in all directions, making these instances better compared to their counter-parts of SA and compass states.

\section{Outlook}\label{sec:outlook}

We now provide a brief discussion of our results with their summarized physical consequences. This comprehensive discussion strives to integrate our findings into current understanding, providing a detailed picture of their effects and contributions to the field.

A four-headed kitten state is considered as main example in our work, which represents a smaller version of a compass state~\cite{Zurek2001}, and our investigation and discussion are centered around this state, which serves as the foundation for our exploration and examination. The compass state exhibits fascinating sub-Planck scale structures and enhanced sensitivity. However, these characteristics are lost when transitioning to a kitten state~\cite{Alexei2006}, which is essentially a smaller version of the cat state, such as the one we presented in Fig.~\ref{fig:fig1}(a). This transition occurs when the macroscopic parameter is reduced, resulting in a transformation from a cat state to a kitten state. Specifically, a bigger compass state exhibits pronounced sub-Planck features, indicating more finer quantum characteristics at scales lower than the Planck length, as discussed in \S\ref{subsec:compass} and then depicted in Fig.~\ref{fig:fig1}. Note that the mean photon number in bigger catlike states is higher~\cite{PhysRevA.99.063813}, indicating enhanced total photon content and intensity. This comparison demonstrates the significant differences in quantum behavior and measurement precision between the larger compass and kitten states, as observed in Fig.~\ref{fig:fig3}.

Photon addition and subtraction operations on squeezed-vacuum states are extremely useful approaches for creating larger cat states~\cite{Dakna1997, Tang201586, Alexei2006, Neergaard2006}. These methods have also been demonstrated experimentally and provide an effective way to generate cat states of larger amplitude~\cite{Alexei2006, Neergaard2006}. In this work, we utilized the kitten version of the compass state and applied photon addition and subtraction operations with different order and magnitudes to construct our variants, as presented in Figs.~\ref{fig:fig8}, \ref{fig:fig9} and \ref{fig:fig10}. We then investigated the phase-space characteristics of these variants to gain insights into their quantum properties. The kitten version of the compass state, as shown in Fig.~\ref{fig:fig1}(a), which does not possess sub-Planckian scale, is now transformed into the states holding sub-Planck structures and demonstrating an enhanced sensitivity, implying the effectiveness of these multiphoton processes, as evident in Fig.~\ref{fig:fig11}. 

When the number of photons added increases, nonclassical features contained by these states are improved, but increasing photon subtraction destroys nonclassical sub-Planck structures in phase space. Note that this only occurs for our SA case; otherwise, when photon subtraction is applied to a state directly (AS cases), the phase space nearly remains unchanged [see Fig.~\ref{fig:fig4}], and the size of the sub-Planck structures of the AS cases stays constant over the variation of the photon subtraction operations, but in this case, under the particular selection of the photon subtraction the isotropy of corresponding sub-Planck structures is improved, as observed in Fig.~\ref{fig:fig10}(a)-\ref{fig:fig10}(c). Furthermore, photon addition raises the average photon count in the coherent states, whereas photon subtraction keeps the photon statistics invariant. This demonstrates that the sequence of photon addition and subtraction operations in our case may have a direct effect on the amount of quantum characteristics present in the states.

Our quantum states exhibit phase-space attributes comparable to the compass state when there is a large number of photon applied to the four-headed kitten state. In addition, the role of photon subtraction in our instances is also interesting as in some cases it greatly enhances the isotropy of these features. Distinct phase-space characteristics between our SA and AS examples are highlighted, namely, that AS cases have smaller sub-Planck structures and so achieve higher sensitivity when compared to their SA counterparts. Our investigation implies that variants of the four-headed kitten state we provided are an appropriate substitute for compass states and may perform better compared to the compass state under optimum conditions.

\section{Conclusion}\label{sec:conc}

We have introduced alternative versions of the compass state, which are obtained by adding (and subtracting) an immense number of photons to a four-headed kitten state, with the option that the order in which these photons are applied to the state also changes and the noncommutativity of the bosonic operators results in two different quantum states with distinct phase-space features. Our outcomes revealed that the multiphoton operations we performed on the four-headed kitten state are quite effective and that these operations have transformed this multicomponent kitten state to other forms of states, which are richer in their nonclassical phase-space features and also exhibit sub-Planck structures. Our investigation also has a close connection with previous experimental studies~\cite{Alexei2006, Neergaard2006}, which demonstrated that photon operations are an effective means of enhancing the phase-space characteristics of quantum states, thereby supporting our findings.

The presence of crucial sub-Planck structures in the present investigation is influenced by the number of photons added or subtracted. Adding photons helps preserve these structures in both scenarios we explored, while subtracting photons typically disrupts them when photon addition is followed by photon subtraction. However, photon subtraction alone has no effect on the sub-Planck structures if applied before photon addition, and in this case (subtraction then addition), specific choices of photon subtraction operations may lead to isotropic versions of sub-Planck structures. These results directly apply to the sensitivity of these states as well. The sensitivity to displacement of the quantum states we proposed exceeds the standard limits, and this enhancement is also controlled by multiphoton operations. Specifically, increasing photon addition enhances sensitivity, while photon subtraction reduces it, in the sequence where photon addition is followed by subtraction. However, in the converse case of photon operations, the sensitivity to displacement remains unchanged over the variation of the photon subtraction, although, interestingly, for this case, improvement in the isotropy of sensitivity is observed for certain photon subtraction choices. The induction of sub-Planck structures in our indicated situations connects them to recent techniques for the development of nonclassical traits in quantum states~\cite{Dakna1997, Tang201586, Alexei2006, Neergaard2006}, and perhaps these techniques further may also be applied in the development of the quantum states we provided.
Future research may inquire how to create our proposed quantum states, which will require a novel and thorough investigation to develop new techniques for their generation. This endeavor may involve formulating innovative strategies and methodologies specifically designed to produce these advanced quantum states.

\section*{Acknowledgment}
This work was supported by the Natural Science Foundation of Jiangsu Province (Grant No. BK20231320) and the National Natural Science Foundation of China (Grant No. 12174157). We would like to thank Tan Hailin for the valuable suggestions and comments on the work.
%\onecolumngrid

\appendix
\section{sub-Planckian scale}\label{appendix:appendixA}

In this appendix, we provide an overview of the fundamental concept of the sub-Planck structure, utilizing graphical illustrations to effectively explore and elucidate this concept. These visual illustrations help in providing a clearer understanding of the intricate details and underlying principles associated with sub-Planck scales.
\begin{figure}[t]
\includegraphics[width=0.48\textwidth]{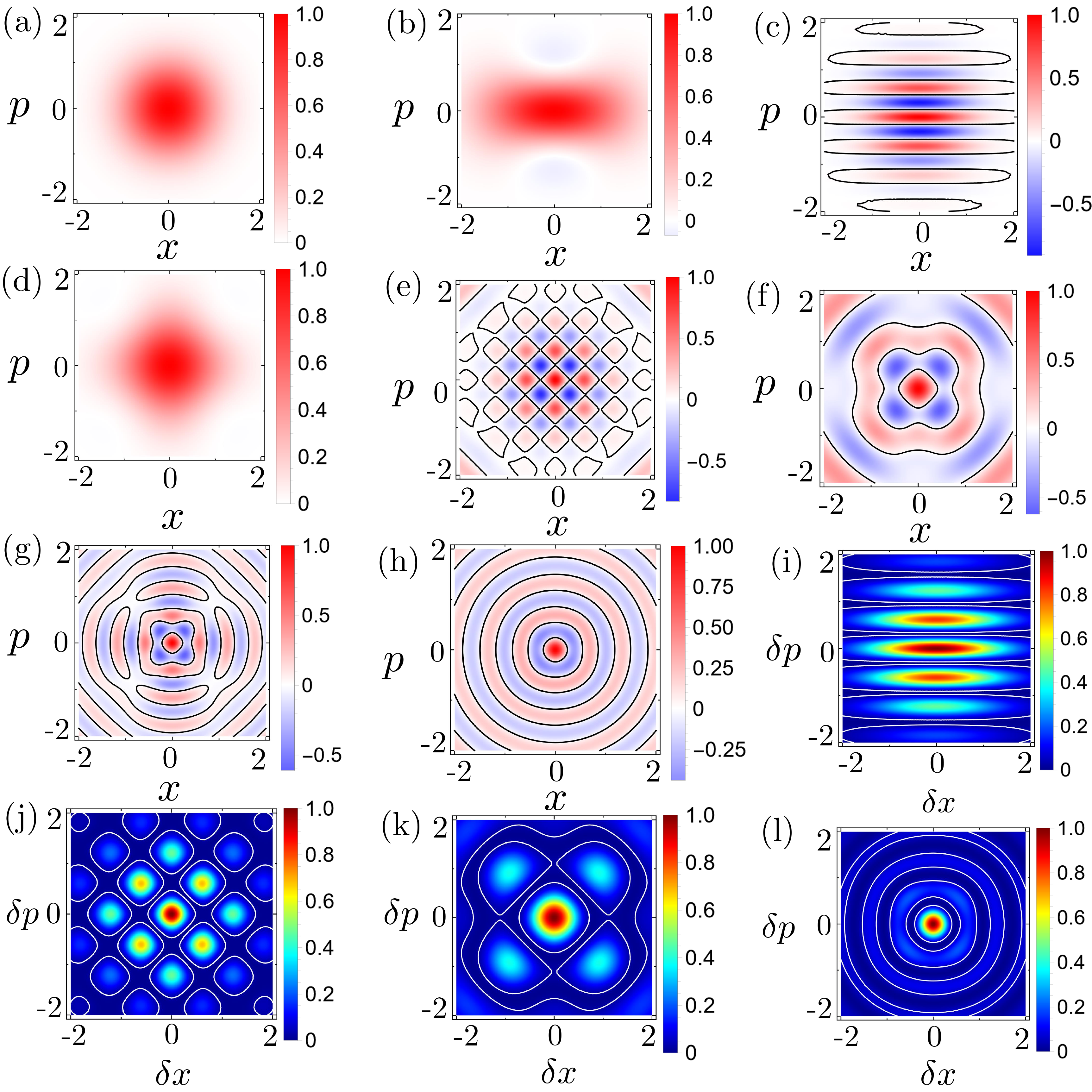}
\caption{[(a)-(h)] Wigner distributions for (a)~vacuum state ($\alpha=0$), (b)~the kitten state with $c_0=1$, (c)~the central interference of the horizontal cat state with $c_0=5$, (d)~the four-headed kitten state of Fig.~\ref{fig:fig1}(a), and the cases corresponding to (e) Fig.~\ref{fig:fig1}(b), (f) Fig.~\ref{fig:fig9}(a), (g)~Fig.~\ref{fig:fig9}(d), and (h)~Fig.~\ref{fig:fig10}(c), which are now plotted with the regions where these distributions get minor amplitudes highlighted by black lines, which are enclosing the corresponding phase-space features. Also shown are the overlap functions for different quantum states of the main text: (i) Overlap between a horizontal cat and its displaced version and [(j)-(l)] overlap functions corresponding to the cases presented in (j)~Fig.~\ref{fig:fig3}(a), (k)~Fig.~\ref{fig:fig11}(a), and (l)~Fig.~\ref{fig:fig11}(i) with white lines marking the zeros of each function.}
\label{fig:figapp}
\end{figure}

\textbf{\small{Planck scale.}} The Wigner function of the coherent state, as given by Eq.~(\ref{eq:wig_coh}), has the phase-space dimension at the Planck scale; that is, the coherent state follows the equality of the Heisenberg uncertainty principle and is considered the minimum uncertainity state~\cite{CarlosNB15}. The Wigner function shown in Fig.~\ref{fig:figapp}(a) for $\alpha = 0$ (a vacuum state) represents a phase-space distribution where the uncertainty in position $\Delta x$ and momentum $\Delta p$ is as small as that set by the minimal standard quantum limit. This means the `volume' of phase space, given by the product $\Delta x \Delta p$, is at its smallest possible value ($\Delta x \Delta p\sim \nicefrac{1}{2}$), in line with the Heisenberg uncertainty principle. Additionally, the spread of the Wigner function in phase space stays the same, meaning its overall shape remains fixed over the variation of associated parameters, staying within the boundaries set by the uncertainty constraint. Hence, simply the phase-space volume of the coherent sets the Planck scale, and the phase-space features of extensions below this norm are simply called sub-Planck scale features~\cite{Zurek2001}.

The smallest phase-space feature associated with the four-headed kitten state investigated in this work also adheres to the Heisenberg uncertainty principle. For example, the case described in the Fig.~\ref{fig:fig1}(a) is replotted with the comparison of the coherent state in Fig.~\ref{fig:figapp}(d) and appears to demonstrate a phase-space volume comparable to that of the coherent state and is consequently considered to be at the Planck scale level.

\textbf{\small{Sub-Planckian scale.}} Figures ~\ref{fig:figapp}(c) and \ref{fig:figapp}(e) depict the central phase-space areas of the cat and compass states, respectively, as explained in the main text. The illustrations emphasize the zeros of the relevant distributions, demonstrating the limitations of corresponding features in the anisotropic or isotropic domains. The black solid lines indicate areas where the Wigner function achieves amplitudes of $10^{-2}$. Upon closer inspection of the central features within this distribution, we notice that they are confined along the black solid lines, which trace out a tile-like pattern in phase space. It is immediately apparent that the size of this central phase-space region is significantly smaller than that of the coherent state depicted in Fig.~\ref{fig:figapp}(a), exhibiting a form of anisotropic sub-Planck structures. Similar sub-Planck structures are present in the situations appearing in Figs.~\ref{fig:figapp}(f), \ref{fig:figapp}(g), and \ref{fig:figapp}(h), which belong to the situations shown in Figs.~\ref{fig:fig9}(a), \ref{fig:fig9}(d), and \ref{fig:fig10}(c), respectively. In Fig.~\ref{fig:fig10}(l) the core sub-Planck structures are contained in a circular region, indicating an isotropic version of the structure.

\textbf{\small{Sensitivities.}} The overlap between the horizontal cat state and its slightly displaced constituent is depicted in Fig.~\ref{fig:figapp}(i). The white lines enclosing structures underline the corresponding zeros of the distributions. The overlap for the compass state case, as shown in Fig.~\ref{fig:fig3}(a), is now represented with the zeros allocated along white lines in Fig.~\ref{fig:figapp}(j). This makes it evident that the central structure has a volume smaller than that of the coherent state, for which $|\delta|<1$ is needed to make the overlap zero. In addition, anisotropic enhancement in the sensitivity is observed, leading to the similar cases shown in Figs.~\ref{fig:figapp}(k) and \ref{fig:figapp}(l), which correspond to the cases illustrated in the main text in Figs.~\ref{fig:fig11}(a) and \ref{fig:fig11}(i) respectively. The case presented in Fig.~\ref{fig:fig11}(i) represents the situation in which isotropic enhancement in the corresponding sensitivity is observed.

\vspace{0.1cm}
%\twocolumngrid
\bibliography{References}
\end{document}